\documentclass[10pt,journal,compsoc]{IEEEtran}

\ifCLASSOPTIONcompsoc
  \usepackage[nocompress]{cite}
\else
  \usepackage{cite}
\fi

\usepackage{amsmath,amssymb,amsfonts}
\usepackage[pdftex]{graphicx}
\usepackage[linesnumbered,lined,ruled]{algorithm2e}
\usepackage{threeparttable}
\usepackage{multirow}
\usepackage{amssymb}
\usepackage[full]{textcomp}
\usepackage{xcolor}
\usepackage{balance}
\usepackage{url}
\usepackage{xspace}
\usepackage[hidelinks]{hyperref}
\usepackage[nameinlink,noabbrev,capitalize]{cleveref}
\usepackage{enumitem}
\usepackage{tcolorbox}
\usepackage{tikz}
\usepackage{ntheorem}
\usepackage{comment}
\usepackage{soul}
\usepackage{xurl}

\definecolor{lightgray}{rgb}{0.95, 0.95, 0.95}
\definecolor{darkgray}{rgb}{0.4, 0.4, 0.4}
\definecolor{purple}{rgb}{0.65, 0.12, 0.82}
\definecolor{editorGray}{rgb}{0.95, 0.95, 0.95}
\definecolor{editorOcher}{rgb}{1, 0.5, 0} %
\definecolor{editorGreen}{rgb}{0, 0.5, 0} %
\definecolor{orange}{rgb}{1,0.45,0.13}
\definecolor{olive}{rgb}{0.17,0.59,0.20}
\definecolor{brown}{rgb}{0.69,0.31,0.31}
\definecolor{purple}{rgb}{0.38,0.18,0.81}
\definecolor{lightblue}{rgb}{0.1,0.57,0.7}
\definecolor{lightred}{rgb}{1,0.4,0.5}
\definecolor{codegreen}{rgb}{0,0.6,0}
\definecolor{codegray}{rgb}{0.5,0.5,0.5}
\definecolor{codepurple}{rgb}{0.58,0,0.82}
\definecolor{backcolour}{rgb}{0.95,0.95,0.92}
\definecolor{airforceblue}{rgb}{0.36, 0.54, 0.66}

\newcommand{\todoviolet}[1]{\todoc{violet}  {[#1]}}

\newcommand{\todoc}[2]{}

\newcommand{\scc}[1]{\todoviolet{scc: #1}}

\newcommand{\revision}[1]{#1}

\makeatletter
\def\namedlabel#1#2{\begingroup
	#2%
	\def\@currentlabel{#2}%
	\phantomsection\label{#1}\endgroup
}
\makeatother

\SetAlCapNameFnt{\footnotesize}
\SetAlCapFnt{\footnotesize}

\newtheorem{property}{Property}

\newtheorem{definition}{Definition}

\newcommand{\srchEvlRecall}{90.19\%\xspace}

\newcommand{\datasetA}{\textbf{$\mathbb{D}^A$}\xspace} %
\newcommand{\datasetS}{\textbf{$\mathbb{D}^S$}\xspace} %
\newcommand{\datasetP}{\textbf{$\mathbb{D}^P$}\xspace}
\newcommand{\dsFromBlock}{13,000,000\xspace}
\newcommand{\dsToBlock}{13,800,000\xspace}
\newcommand{\dsTotalBlocks}{800,000\xspace}
\newcommand{\dsTotalAttacks}{188,700\xspace}

\newcommand{\benchmarkA}{\textbf{$\mathbb{B}^A$}\xspace}
\newcommand{\benchmarkTotalAttacks}{513\xspace}
\newcommand{\benchmarkTotalContracts}{235\xspace}

\newtheorem{finding}{Finding}
\newtheorem*{implication}{Implication}

\newcommand{\rqOneTotalSamples}{383\xspace}

\newcommand{\totalTools}{seven\xspace}

\newcommand{\minSampleSize}{200\xspace}

\usepackage{listings, xcolor}
\usepackage[T1]{fontenc}
\usepackage{lipsum} %

\definecolor{verylightgray}{rgb}{.97,.97,.97}

\lstdefinelanguage{Solidity}{
	keywords=[1]{anonymous, assembly, assert, balance, break, call, callcode, case, catch, class, constant, continue, constructor, contract, debugger, default, delegatecall, delete, do, else, emit, event, experimental, export, external, false, finally, for, function, gas, if, implements, import, in, indexed, instanceof, interface, internal, is, length, library, log0, log1, log2, log3, log4, memory, modifier, new, payable, pragma, private, protected, public, pure, push, require, return, returns, revert, selfdestruct, send, solidity, storage, struct, suicide, super, switch, then, this, throw, transfer, true, try, typeof, using, value, view, while, with, addmod, ecrecover, keccak256, mulmod, ripemd160, sha256, sha3}, %
	keywordstyle=[1]\color{blue}\bfseries,
	keywords=[2]{address, bool, byte, bytes, bytes1, bytes2, bytes3, bytes4, bytes5, bytes6, bytes7, bytes8, bytes9, bytes10, bytes11, bytes12, bytes13, bytes14, bytes15, bytes16, bytes17, bytes18, bytes19, bytes20, bytes21, bytes22, bytes23, bytes24, bytes25, bytes26, bytes27, bytes28, bytes29, bytes30, bytes31, bytes32, enum, int, int8, int16, int24, int32, int40, int48, int56, int64, int72, int80, int88, int96, int104, int112, int120, int128, int136, int144, int152, int160, int168, int176, int184, int192, int200, int208, int216, int224, int232, int240, int248, int256, mapping, string, uint, uint8, uint16, uint24, uint32, uint40, uint48, uint56, uint64, uint72, uint80, uint88, uint96, uint104, uint112, uint120, uint128, uint136, uint144, uint152, uint160, uint168, uint176, uint184, uint192, uint200, uint208, uint216, uint224, uint232, uint240, uint248, uint256, var, void, ether, finney, szabo, wei, days, hours, minutes, seconds, weeks, years},	%
	keywordstyle=[2]\color{teal}\bfseries,
	keywords=[3]{block, blockhash, coinbase, difficulty, gaslimit, number, timestamp, msg, data, gas, sender, sig, value, now, tx, gasprice, origin},	%
	keywordstyle=[3]\color{violet}\bfseries,
	identifierstyle=\color{black},
	sensitive=false,
	comment=[l]{//},
	morecomment=[s]{/*}{*/},
	commentstyle=\color{gray}\ttfamily,
	stringstyle=\color{red}\ttfamily,
	morestring=[b]',
	morestring=[b]"
}

\begin{document}

\title{Combatting Front-Running in Smart Contracts: Attack Mining, Benchmark Construction and Vulnerability Detector Evaluation}

\author{Wuqi~Zhang,~\IEEEmembership{Graduate Student Member,~IEEE,}
	Lili~Wei,
	Shing-Chi~Cheung,~\IEEEmembership{Fellow,~IEEE,}
	Yepang~Liu,~\IEEEmembership{Member,~IEEE,}
	Shuqing~Li,
	Lu~Liu,
	and~Michael~R.~Lyu,~\IEEEmembership{Fellow,~IEEE}
	\IEEEcompsocitemizethanks{
		\IEEEcompsocthanksitem Wuqi Zhang, Shing-Chi Cheung, and Lu Liu are with the Department
		of Computer Science and Engineering, The Hong Kong University of Science and Technology, Hong Kong, China.\protect\\
		E-mail: wuqi.zhang@connect.ust.hk, scc@cse.ust.hk, \\
		lliubf@connect.ust.hk
		\IEEEcompsocthanksitem Lili Wei is with the Department of Electrical and Computer Engineering, McGill University, Montreal, Quebec, Canada.\protect\\
		E-mail: lili.wei@mcgill.ca
		\IEEEcompsocthanksitem Yepang Liu is with the Department of Computer Science and Engineering, and the Research Institute of Trustworthy Autonoumous Systems, Southern University of Science and Technology, Shenzhen, Guangdong, China.\protect\\
		E-mail: liuyp1@sustech.edu.cn
		\IEEEcompsocthanksitem Shuqing Li and Michael R. Lyu are with the Department of Computer Science and Engineering, The Chinese University of Hong Kong, Hong Kong, China.\protect\\
		E-mail: sqli21@cse.cuhk.edu.hk, lyu@cse.cuhk.edu.hk
		\IEEEcompsocthanksitem Shing-Chi Cheung is the corresponding author.
	}%
}

\ifCLASSOPTIONpeerreview
	\markboth{Journal of \LaTeX\ Class Files,~Vol.~14, No.~8, August~2015}%
	{Shell \MakeLowercase{\textit{et al.}}: Bare Demo of IEEEtran.cls for Computer Society Journals}
\fi

\IEEEtitleabstractindextext{%
	\begin{abstract}
		Front-running attacks have been a major concern on the blockchain.
		Attackers launch front-running attacks by inserting additional transactions before upcoming victim transactions to manipulate victim transaction executions and make profits.
		Recent studies have shown that front-running attacks are prevalent on the Ethereum blockchain and have caused millions of US dollars loss.
		It is the vulnerabilities in smart contracts, which are blockchain programs invoked by transactions, that enable the front-running attack opportunities.
		Although techniques to detect front-running vulnerabilities have been proposed, their performance on real-world vulnerable contracts is unclear.
		There is no large-scale benchmark based on real attacks to evaluate their capabilities.
		We make four contributions in this paper.
		First, we design an effective algorithm to mine real-world attacks in the blockchain history.
		The evaluation shows that our mining algorithm is more effective and comprehensive, achieving higher recall in finding real attacks than the previous study.
		Second, we propose an automated and scalable vulnerability localization approach to localize code snippets in smart contracts that enable front-running attacks.
		The evaluation also shows that our localization approaches are effective in achieving higher precision in pinpointing vulnerabilities compared to the baseline technique.
		Third, we build a benchmark consisting of \benchmarkTotalAttacks real-world attacks with vulnerable code labeled in \benchmarkTotalContracts distinct smart contracts, which is useful to help understand the nature of front-running attacks, vulnerabilities in smart contracts, and evaluate vulnerability detection techniques.
		Last but not least, we conduct an empirical evaluation of \totalTools state-of-the-art vulnerability detection techniques on our benchmark.
		The evaluation experiment reveals the inadequacy of existing techniques in detecting front-running vulnerabilities, with a low recall of $\leq$~6.04\%.
		Our further analysis identifies four common limitations in existing techniques: lack of support for inter-contract analysis, inefficient constraint solving for cryptographic operations, improper vulnerability patterns, and lack of token support.
	\end{abstract}

	}

\maketitle
\IEEEdisplaynontitleabstractindextext
\IEEEpeerreviewmaketitle

\IEEEraisesectionheading{\section{Introduction}\label{sec:introduction}}

\IEEEPARstart{F}ront-running \cite{wikipediaFrontRunning2022x} attacks in financial markets refer to the practice of leveraging the knowledge of future transactions and trading before them to make profits.
Front-running attacks also occur in blockchain systems like Ethereum~\cite{woodEthereumSecureDecentralised2020}, where transactions are published before execution.
Upcoming transactions are available to all blockchain users, including potential attackers.
By adjusting transaction execution orders with miners~\cite{daianFlashBoysFrontrunning2019}, malicious attackers can attack victims by executing transactions before victim ones so that the victim transactions would be executed on different blockchain states from what was expected.
As a result, the attackers can make profits from the attack and cause financial losses to the victims.
Smart contracts, the programs invoked by transactions to perform actions on the blockchain, could make front-running profitable for attackers.
Fig.~\ref{lst:transfer_manager_contract} shows an example smart contract vulnerable to front-running attacks.
A relayer (\texttt{msg.sender}) provides a relay service for off-chain users, who may not have enough Ethers to pay for transaction fees, to perform on-chain operations.
The relayer calls function \texttt{relayOperation} to execute user operations (Line~\ref{lst:transfer_manager_contract:relay}) and charges ERC20~\cite{vogelstellerEIP20TokenStandard2023} tokens (line~\ref{lst:transfer_manager_contract:fee}) as the profits of providing the service.
Front-running attacks may occur since the users' operations and signatures used to invoke this contract are publicly available once the relayer submits the transaction.
One attacker can invoke this contract before the relayer and take the profits, which should have been given to the relayer.
As a consequence, the relayer's transaction fails since each user operation can only be executed once (line~\ref{lst:transfer_manager_contract:uniq}).
The profits are taken by the attacker even if it is the relayer who makes efforts to provide the relay service (e.g.,~maintaining an easy-to-use interface like Web Apps).

\begin{figure}
	\begin{lstlisting}[language=Solidity]
contract TransferManager {
  function relayOperation(
    ERC20 token, address user,
    bytes operation, bytes signature
  ) {
    if(verifySignature(
      user, operation, signature
    )) {/*\label{lst:transfer_manager_contract:verify}*/
      require(/*\label{lst:transfer_manager_contract:uniq}*/
        checkUniqueness(
          user, operations, signature
        ),
        "RM: Duplicate request"
      );
      executeOperation(operation);  /*\label{lst:transfer_manager_contract:relay}*/
      // tarnsfer relay fee to relayer
      uint relayFee = getFee();
      token.transferFrom(/*\label{lst:transfer_manager_contract:fee}*/
        user, msg.sender, relayFee
      );
}}}
    \end{lstlisting}
	\caption{Simplified version of TransferManager contract~\cite{etherscan.ioTransferManagerEtherscan2023} on Ethereum.
		Attackers attack by invoking \texttt{relayOperation} function before victim transactions.}
	\label{lst:transfer_manager_contract}
\end{figure}

Recent studies have revealed the prevalence and severity of front-running attacks on Ethereum by conducting measurement studies~\cite{eskandariSoKTransparentDishonesty2020,torresFrontrunnerJonesRaiders2021,daianFlashBoysFrontrunning2019}.
Torres et al.~\cite{torresFrontrunnerJonesRaiders2021} found that front-running attacks are prevalent on the Ethereum blockchain and have caused a total loss of over 18.41M USD.
Daian et al.~\cite{daianFlashBoysFrontrunning2019} pointed out that front-running attacks also pose a major threat to the ecosystem of blockchain since the profits of front-running attacks could be larger than the cost of forking the blockchain.
Attackers could share part of the profits to attract miners to fork the blockchain.
Given the great impact, many researchers aimed to curb front-running in smart contracts.
Vulnerabilities under front-running attacks have been named in different ways, such as transaction order dependency~\cite{luuMakingSmartContracts2016}, event ordering bugs~\cite{kolluriExploitingLawsOrder2019}, and state inconsistency bugs~\cite{boseSAILFISHVettingSmart2022}.
In this paper, we refer to them generally as front-running vulnerabilities.
Various techniques~\cite{luuMakingSmartContracts2016,tsankovSecurifyPracticalSecurity2018,muellerSmashingEthereumSmart2018,velosoConkasModularStatic,kolluriExploitingLawsOrder2019,boseSAILFISHVettingSmart2022} have been proposed to detect such vulnerabilities in smart contracts.
\revision{
	However, these techniques are usually evaluated in terms of detection precision on real-world smart contracts.
	The recall rate is hard to evaluate due to the lack of ground truth.
	Ghaleb and Pattabiraman~\cite{ghalebHowEffectiveAre2020} propose to inject vulnerable code into contracts to evaluate the vulnerability detection recall of contract analyzers.
	However, the front-running vulnerabilities injected are limited to a few rigid code snippets and cannot represent vulnerabilities in real-world contracts.
	There still lacks a large-scale systematic study to evaluate and understand the performance of these detection techniques on real-world front-running vulnerabilities.
}

To address the research gap and evaluate the detection techniques, the major challenge is how to build a large-scale and representative benchmark with ground truth of vulnerabilities.
Existing studies cannot fill the gap due to two major limitations.
On the one hand, the existing benchmark is neither large-scale nor contains representative contracts to evaluate front-running vulnerability detection tools.
An existing empirical study~\cite{durieuxEmpiricalReviewAutomated2020} offers a benchmark with only four simple vulnerable contracts, with 33.75 lines of code in each on average.
On the other hand, datasets of real-world attacks built by the previous measurement studies~\cite{torresFrontrunnerJonesRaiders2021,daianFlashBoysFrontrunning2019} cannot be used as benchmarks for the evaluation of vulnerability detection techniques.
There are two main reasons.
First, the measurement studies rely on several predefined patterns and heuristic rules to match attacks in history, which may not be comprehensive and potentially miss many attacks.
As to be shown in Section~\ref{sec:search:evaluation}, the approach proposed by our study is able to identify 24.42x time attacks than the existing dataset~\cite{torresFrontrunnerJonesRaiders2021} with 98.69\% precision, indicating that many attacks are actually missed by the existing work.
Second and most importantly, none of the measurement studies can localize vulnerabilities in smart contracts.
Only attacks, each consisting of several transactions, are identified, while it is still unknown which code in the underlying smart contracts enables the possibility of front-running attacks.
\revision{
	To tackle the limitations of previous studies, we first propose a general attack model and mine historical front-running attacks using the model.
	The mined attacks serve as ground truth, from which we propose a novel technique to localize front-running vulnerabilities in smart contracts and build a benchmark.
	To demonstrate the usefulness of our approach, we systematically evaluate \totalTools state-of-the-art tools and investigate the limitations in their techniques.
}

Our attack mining algorithm enumerates all transactions in history with efficient pruning strategies and a generic attack model.
Previous works~\cite{torresFrontrunnerJonesRaiders2021,daianFlashBoysFrontrunning2019} rely on a limited number of predefined patterns to find historical attacks, which can miss many attacks.
Our evaluation results show that our algorithm can achieve \srchEvlRecall recall on a baseline dataset~\cite{torresFrontrunnerJonesRaiders2021} and find 24.42x more attacks than the state-of-the-art technique.
It also has precision as high as 98.69\% since we strictly follow the definition of front-running in the attack model.
We mine historical attacks in the latest \dsTotalBlocks blocks on the Ethereum mainnet and collect \dsTotalAttacks attacks in total.

With a large-scale dataset of real attacks, we localize vulnerable code in smart contracts.
For each attack, we consider the blockchain shared data manipulated by the attacker as taint sources and perform dynamic taint analysis with the victim transaction.
We consider the program location where victim profits are directly affected as the taint sink.
Then we mark the contract code executed along the taint flow trace from source to sink as vulnerable.
Our manual analysis among three authors on a sample of attacks shows that the code localized by our approach can cover the exploited vulnerable contract logic in all attacks.
In addition, we also find that our approach is precise and marks 99.66\% less code than the baseline.
In the end, we build a benchmark with \benchmarkTotalAttacks real-world attacks and identify the vulnerabilities in \benchmarkTotalContracts contracts whose source code is available.

Based on the benchmark, we perform an empirical study to evaluate existing techniques.
We aimed to answer the following research questions.
\begin{itemize}[leftmargin=*]
	\item  How many vulnerabilities can existing tools detect in our benchmark?
	\item What are the limitations of existing tools in detecting front-running vulnerabilities?
\end{itemize}
We conduct a systematic literature review on state-of-the-art works and select \totalTools tools that implement techniques supporting front-running vulnerability detection.
We use these tools to detect vulnerabilities in our benchmark.
Our results show that existing tools have poor performance and can only detect vulnerabilities exploited by at most 6.04\% attacks.
We then investigate the limitations of the underlying techniques of each tool through manual analysis of samples of missed vulnerabilities.
Our major findings include:
\begin{itemize}[leftmargin=*]
	\item Existing techniques can hardly perform precise inter-contract analysis, failing to capture many vulnerabilities involving cross-contract invocations.
	\item The wide use of cryptographic operations in contracts makes it difficult to generate concrete transactions using SMT solvers, limiting the capability of the techniques in exploring transaction executions.
	\item Vulnerability detection patterns of existing techniques are weak in capturing many front-running vulnerabilities.
	\item Many vulnerabilities are missed due to the negligence of profit-making in tokens instead of Ethers.
\end{itemize}

\revision{
	To sum up, we make the following contributions in this work.
	\begin{itemize}[leftmargin=*]
		\item We design an effective algorithm to comprehensively mine front-running attacks in the Ethereum transaction history.
		\item We propose a novel approach to automatically localize vulnerable code from a historical attack.
		\item We build a benchmark consisting of \benchmarkTotalAttacks real-world attacks with vulnerable code localized in \benchmarkTotalContracts distinct contracts, which can be used to evaluate existing vulnerability detection techniques.
		\item We conduct an empirical evaluation on \totalTools state-of-the-art vulnerability detection tools and find that none of them are effective, only with a low recall $\leq 6.04\%$.
		      We investigate and summarize four common limitations of the techniques, providing insights for future improvements of front-running vulnerability detection.
	\end{itemize}
}

The implementation of our attack mining and vulnerability localization, our benchmark, as well as the results of our evaluations of existing tools, are made publicly available on GitHub:
\url{https://github.com/Troublor/erebus-redgiant}.

\revision{
	The following of this paper is organized as follows:
	Section~\ref{sec:background} introduces the background knowledge of front-running on the blockchain.
	Section~\ref{sec:literature} discusses literature related to front-running attacks and vulnerabilities in smart contracts.
	Section~\ref{sec:search} aims to build a high-quality dataset of historical front-running attacks.
	Section~\ref{sec:locate} is meant to build a front-running vulnerability benchmark by localizing vulnerabilities from the attack dataset built in Section~\ref{sec:search}.
	Section~\ref{sec:evaluation} evaluates state-of-the-art contract analyzers using the benchmark built in Section~\ref{sec:locate} and discusses the limitation of these analyzers.
}

\section{Background}
\label{sec:background}

This section introduces the background knowledge of the Ethereum blockchain and front-running attacks.
We base our presentation on Ethereum since it is the most popular blockchain that supports Turing-complete smart contracts.
In this paper, blockchain refers to the Ethereum blockchain unless otherwise specified.

\subsection{Ethereum State Transition Model}

Ethereum blockchain can be considered a state machine~\cite{woodEthereumSecureDecentralised2020}.
State transitions occur when transactions get executed in new blocks mined by miners.
A global state called \textit{world state} is maintained by Ethereum.
The blockchain world state comprises account cryptocurrency balances (in Ethers), smart contract code, and key-value mapping storage for each smart contract.
Every executed transaction modifies the world state by performing a simple cryptocurrency transfer or invoking a smart contract, which is the program stored on the blockchain specifying the logic of world state modification.
In order to achieve the consensus of state transitions across all blockchain miners, the execution of a transaction is deterministic given a pre-execution world state, and transactions are executed sequentially according to an order determined by miners.

\subsection{Transaction Order in Blocks}
\label{background:transaction-order}
The order transactions executed in each block are determined by miners to enlarge their profits.
Miners make profits by charging execution fees for each transaction~\cite{ethereumfoundationEthereumOrg2021}.
The execution fee is calculated by the multiplication of gas, which measures the amount of computing resource consumed in the execution, and the gas price.
To maximize profits, miners usually prioritize transactions that specify higher gas prices~\cite{daianFlashBoysFrontrunning2019}.
Users of Ethereum can set a relatively higher gas price to prioritize the transaction execution.
Note that executing transactions in descending order of gas price is not a must.
Miners are free to order transactions to their interest.

\subsection{Front-Running Attacks}
\label{sec:background:attacks}
Front-running attacks have been clearly defined by several Ethereum vulnerability taxonomies~\cite{smartcontractsecuritySWCRegistrySmart2021,nccgroupDASPTOP102021}.
Attackers leverage the information revealed by future transactions, execute attack transactions in advance to make profits, and result in unexpected behavior in the victim transactions.
On Ethereum, before execution, pending transactions are stored in a pool, broadcast to all miners, and known to attackers.
Attackers can easily obtain the information revealed by the pending transactions and construct attack transactions to perform front-running attacks.
To execute attack transactions before the victim, the attacker can either set a higher gas price or mine blocks themselves, as mentioned in Section~\ref{background:transaction-order}.
As a result, the victim transaction has an execution outcome different from that without the attack transaction executed, causing loss to the victim transaction user.  %

\begin{figure}
	\begin{lstlisting}[language=Solidity]
contract Swap {
  Pair pair;
  function swap(uint amount0) public {/*\label{lst:uniswap:swap}*/
	// Charge constant swap fee
    amount0 = amount0 - 100gwei; // swap fees /*\label{lst:uniswap:cut-fee}*/
    pair.token0.transferFrom( /*\label{lst:uniswap:fee}*/
        msg.sender, this, 100gwei
    );
	// Calculate the amount of token1 swapped to
    uint amount1 = pair.reserve0*pair.reserve1/(pair.reserve0+amount0)-pair.reserve1; /*\label{lst:uniswap:calculation}*/
	// Log the swap event
    logSwap(msg.sender, amount0, amount1); /*\label{lst:uniswap:log}*/
	// Swap the tokens
    pair.doSwap(amount0, amount1); /*\label{lst:uniswap:external-call}*/
  }
  function logSwap(
    address u, uint amountIn, uint amountOut
  ) {
    // log the swap event
    ...
  }
}
contract Pair {
  uint public reserve0, reserve1;
  ERC20 public token0, token1;
  function doSwap(uint amount0, uint amount1) public {
	// Update the reserve of the token pair
    reserve0 += amount0; reserve1 -= amount1; /*\label{lst:uniswap:alteration}*/
	// Transfer the swapped tokens
    token0.transferFrom(tx.origin, this, amount0);/*\label{lst:uniswap:transfer-token-in}*/
    token1.transferFrom(this, msg.sender, amount1); /*\label{lst:uniswap:profit}*/
  }
}
    \end{lstlisting}
	\caption{Simplified version of UniswapV2~\cite{uniswapDecentralizedTradingProtocol2021} contract. Attackers invoke function \texttt{swap} before victims. Attackers can buy \texttt{token1} with \texttt{token0} at a lower price, and sell \texttt{token1} afterward at a higher price to make arbitrage.}
	\label{lst:uniswap}
\end{figure}

Front-running can occur in traditional financial markets.
For instance, in foreign exchange markets~\cite{evansFrontRunningCollusionForex2019}, malicious traders can leverage internal information about upcoming large EUR purchase orders, buy EUR using USD in advance at a lower price, and sell them back to USD afterward at a much higher price.
As a result, the upcoming (victim) transaction buys EUR at a higher price, while the malicious traders (attackers) obtain profits from the price difference.
Such markets are also implemented on the blockchain.
\revision{
	Fig.~\ref{lst:uniswap} shows the simplified logic of a popular token exchange market, Uniswap~\cite{uniswapDecentralizedTradingProtocol2021}, which contains front-running vulnerabilities and enables attacks similar to those in foreign exchange markets.
	The contract \texttt{Pair} holds the reserves of two tokens to swap (\texttt{token0} and \texttt{token1}) in variables \texttt{reserve0} and \texttt{reserve1}, respectively.
	The exchange rate between these two tokens is determined by the ratio of reserves held by contract \texttt{Pair}.
	The \texttt{swap} function swaps the given amount of \texttt{token0} (\texttt{amount0}) to \texttt{token1}.
	The swapped \texttt{amount1} is calculated using reserves of contract \texttt{Pair} at line~\ref{lst:uniswap:calculation}.
	The victim transaction swapping \texttt{token0} for \texttt{token1} can be attacked if the attacker invokes function \texttt{swap} in advance.
	The attacker's transaction will modify the values in variables \texttt{reserve0} and \texttt{reserve1} at line~\ref{lst:uniswap:alteration}, changing the ratio of two tokens' reserves.
	As a result, the victim receives less \texttt{token1}, according to the calculation of \texttt{amount1} at line~\ref{lst:uniswap:calculation}, i.e.,~the victim buys \texttt{token1} at a higher price.
	The attacker can later sell \texttt{token1} at a much higher price after the victim transaction to make profits.
}

Front-running is illegal in traditional markets regulated by the government.
However, there is no similar governance on the blockchain.
Attacks are much easier to launch since malicious users can easily know upcoming transactions from the public pool of pending transactions.
Inserting attack transactions before victims is possible since the execution orders are determined by miners without any restrictions.
Therefore, front-running attacks are prevalent on the blockchain and cause much damage~\cite{torresFrontrunnerJonesRaiders2021}.

\section{Literature Review}
\label{sec:literature}

\subsection{Smart Contract Vulnerability and Detection}
\label{sec:review:techniques}

Researchers have identified many different types of vulnerabilities in smart contracts~\cite{smartcontractsecuritySWCRegistrySmart2021}, including integer overflow/underflow, reentrancy, denial of service, and etc.
Various techniques have been proposed to detect these vulnerabilities~\cite{luuMakingSmartContracts2016,tsankovSecurifyPracticalSecurity2018,kolluriExploitingLawsOrder2019,jiangContractFuzzerFuzzingSmart2018,xueCrosscontractStaticAnalysis2020,wangDetectingNondeterministicPayment2019,heLearningFuzzSymbolic2019,grechMadMaxSurvivingOutofgas2018,boseSAILFISHVettingSmart2022,nguyenSFuzzEfficientAdaptive2020,zhuangSmartContractVulnerability2020,soSmarTestEffectivelyHunting2021,choiSMARTIANEnhancingSmart2021,wustholzTargetedGreyboxFuzzing2020,kruppTeEtherGnawingEthereum2018,kalraZEUSAnalyzingSafety2018,mossbergManticoreUserFriendlySymbolic2019,torresOsirisHuntingInteger2018,wustholzHarveyGreyboxFuzzer2020,griecoEchidnaEffectiveUsable2020,liuReGuardFindingReentrancy2018,liuSgramSemanticawareSecurity2018,tikhomirovSmartCheckStaticAnalysis2018,feistSlitherStaticAnalysis2019,muellerSmashingEthereumSmart2018,wangVultronCatchingVulnerable2019}.
Among them, we focus on those techniques capable of detecting front-running vulnerabilities in smart contracts.
Such vulnerability captures the key of front-running attacks: transaction order could influence the execution results.
The authors then proposed Oyente~\cite{luuMakingSmartContracts2016} the first one detecting front-running vulnerabilities, bycheckingE whether there are different ther transfer flows in different execution paths using symbolic execution~\cite{cowardSymbolicExecutionTesting1991}.
Following Oyente, many other vulnerability analyzers support transaction order dependency detection using various techniques.
Ethracer~\cite{kolluriExploitingLawsOrder2019} adopts dynamic symbolic execution to generate concrete transactions and checks whether the resulting blockchain world state is sensitive to the execution orders of these transactions.
Mythril~\cite{muellerSmashingEthereumSmart2018} and Conkas~\cite{velosoConkas2022} leverage symbolic execution and static taint analysis~\cite{schwartzAllYouEver2010} to detect front-running vulnerabilities by checking whether there are feasible execution paths where Ether transfers are affected by taint sources, which are contract storage that another transaction can modify.
Securify~\cite{tsankovSecurifyPracticalSecurity2018} uses abstract intepretation~\cite{cousotAbstractInterpretation1996} to match contract with security property patterns, i.e., the receiver, amount, and path conditions of Ether transfers should not depend on variables that another transaction can manipulate.
Similarly, Sailfish~\cite{boseSAILFISHVettingSmart2022} builds the smart contract state dependency graph, summarizing the read-write dependencies between different public functions, which different transactions invoke.
Then, the same security patterns of Securify are applied to the state dependency graph to detect vulnerabilities.

\subsection{Vulnerability Empirical Studies and Benchmarks}
\revision{
	In addition to various vulnerability detection techniques proposed, researchers have made efforts to evaluate the capability of these techniques from different aspects.
	Durieux et al.~\cite{durieuxEmpiricalReviewAutomated2020} collect two datasets of smart contracts.
	One dataset has vulnerabilities labeled, which can be used to evaluate the vulnerability detection recall rate.
	However, this dataset is small-scale, only containing four simple contracts suffering from front-running vulnerabilities, with 33.75 lines of code on average.
	The other dataset built by Durieux et al.~\cite{durieuxEmpiricalReviewAutomated2020} is large-scale, containing 47,518 contracts.
	However, this dataset does not have the ground truth of the vulnerabilities.
	Ghaleb and Pattabiraman~\cite{ghalebHowEffectiveAre2020} propose to inject vulnerabilities into real-world smart contracts to obtain the ground truth of vulnerabilities for detection technique evaluation.
	However, However, their approach can only inject a few rigid vulnerable code snippets in smart contracts, which may not be the real vulnerable code exploited in real-world attacks.
	In addition, in our study, we find that sometimes the front-running vulnerability resides in the business logic of interaction among multiple contracts, e.g.,~the example in Fig.~\ref{lst:uniswap}.
	Such cross-contract vulnerability is closely coupled with the semantics of the underlying contracts and can not be injected with the proposed injection approach.
	Our study is meant to build a large-scale benchmark with vulnerability ground truth in real-world contracts, which significantly differs from the previous studies.
	As shown in Section~\ref{sec:evaluation}, our benchmark is able to reveal the performance of various contract analyzers in real-world contracts.
	Perez and Livshits~\cite{perezSmartContractVulnerabilities2021} conduct an empirical study on the contracts that are reported as vulnerable by various analyzers.
	Their results show that most front-running vulnerabilities reported by analyzers are benign and have not been exploited in history.
	Their study focuses on the exploitability of the vulnerabilities reported by analyzers and only mines attacks on those reported contracts.
	However, as shown in our study (Section~\ref{sec:evaluation}), state-of-the-art analyzers miss a large number of vulnerable contracts that suffer from front-running attacks.
	As a result, most of the front-running attacks are not included in the dataset collected by Perez and Livshits~\cite{perezSmartContractVulnerabilities2021}.
	Besides, their study has different objectives from ours.
	Their objective is to study the precision of analyzers in terms of exploitability while ours is meant to build a benchmark and evaluate the detection recall rate in real-world contracts.
}

\subsection{Real-World Attacks and Measurement Study}
\label{sec:review:attacks}

Although researchers have identified front-running attacks for years, such attacks have always been prevalent in real-world smart contracts.
Daian et al.~\cite{daianFlashBoysFrontrunning2019} analyzed the transaction traffic on the Ethereum blockchain, showing that many arbitrage bots are competing with each other to perform front-running attacks on transactions submitted by ordinary users automatically.
Eskandari et al.~\cite{eskandariSoKTransparentDishonesty2020} conducted a case study on four categories of smart contracts and found that front-running attacks could happen in contracts designed for cryptocurrency exchange markets, crypto-collectible games, gambling, and name services.
From the case study, the authors identified three attack patterns, i.e., displacement, insertion, and suppression.
Displacement attacks usually observe the input of the victim transaction, invoke the contract in advance as the victim would do, and obtain any profit that would be given to the victim transaction sender.
The example contract in Fig.~\ref{lst:transfer_manager_contract} is vulnerable to displacement attacks.
An insertion attack is performed by inserting a transaction before the victim transaction, altering the state that the victim transaction will executethe  based on, and then executing another transaction after the victim to collect profits.
The example attack in financial market mentioned in Section~\ref{sec:background:attacks} is a typical insertion attack.
A suppression attack is meant to attack time-sensitive transactions by filling the current block and delaying the victim transaction.
Based on the findings from Eskandari et al., Torres et al.~\cite{torresFrontrunnerJonesRaiders2021} took the first step to measure the real-world front-running attacks on Ethereum.
They identified around 200 thousand attacks from the blockchain transaction history and found that displacement and insertion attacks take the majority, obtaining an accumulated profit of 18.41M USD.
Qin et al.~\cite{qinQuantifyingBlockchainExtractable2022} also conducted a similar measurement study on the Ethereum blockchain, also showing that front-running attacks are prevalent and causing considerable financial loss.

\section{Attack Mining and Dataset}
\label{sec:search}
We aim to build a benchmark of vulnerable contracts from real-world front-running attacks.
However, it is non-trivial to mine historical attacks given the large search space, and there exists no generic attack model to identify front-running attacks.
This section introduces our attack model, based on which we propose an algorithm to effectively and comprehensively mine attacks in the blockchain transaction history.

\subsection{Attack Model}
\label{sec:search:model}
We model one front-running attack in blockchain transaction history with a tuple of transactions: $\langle T_a, T_v, T_a^p \rangle$, where $T_v$ is the victim transaction being attacked, and $T_a$ and $T_a^p$ are transactions from the attacker.
We define two transaction execution scenarios:
\begin{definition}[Attack Scenario]
	The tuple of transactions are executed in the order $T_a \rightarrow T_v \rightarrow T_a^p$.
\end{definition}
\begin{definition}[Attack-Free Scenario]
	The tuple of transactions are executed in the order $T_v \rightarrow T_a \rightarrow T_a^p$.
\end{definition}
The attack scenario refers to the execution order in the blockchain history where the attack occurred.
The attack-free scenario refers to the execution order without interference from attackers, which was intended by the victim.

We consider $A = \langle T_a, T_v, T_a^p \rangle$ as a front-running attack if it satisfies two properties:
\begin{property}[Attacker Gain]
	\label{prop:attacker-gain}
	The attacker obtains financial gain in the attack scenario compared with the attack-free scenario.
\end{property}
\begin{property}[Victim Loss]
	\label{prop:victim-loss}
	The victim suffers from financial loss in the attack scenario compared with the attack-free scenario.
\end{property}

\revision{
	The financial gain and loss of attackers and victims are measured by the amount of cryptocurrency and tokens that attackers and victims receive in the transactions.
	We consider Ether, the native cryptocurrency on Ethereum, as well as four popular token standards as quantitative financial profits, namely ERC20~\cite{vogelstellerEIP20TokenStandard2023}, ERC721~\cite{entrikenEIP721NonFungibleToken2023}, ERC777~\cite{dafflonEIP777TokenStandard2023}, and ERC1155~\cite{radomskiEIP1155MultiToken2023}, which are all the token standards\footnote{We do not include ERC4626~\cite{santoroERC4626TokenizedVaults2023} here since it is an extension of ERC20. Supporting ERC20 will support ERC4626 tokens intrinsically.} listed in the official documentation of Ethereum~\cite{ethereumfoundationTokenStandards2023}.
	It is possible that in some specific contracts, the attacker may target other forms of assets besides the standard cryptocurrency and tokens.
	We do not include non-standard assets since our attack model is designed to be general and not limited to specific contracts.
	If necessary, our attack model can be easily extended to support other non-standard forms of assets.
	One only needs to define profit gain or loss on the non-standard asset, and our approach will work seamlessly.
}

The intuition of our attack model is that the attacker should steal benefits from the victim by inserting $T_a$ and manipulating the world state on which $T_v$ executes.
The Attacker Gain property specifies that the attacker benefits from front-running victim transactions.
\revision{If not, the attacker has no incentive to front-run the victim transaction since prioritizing the execution of transactions requires extra costs (Section \ref{background:transaction-order}).}
The Victim Loss property specifies that the victim is harmed due to the attack transaction executed in the front.
The Victim Loss property is designed to exclude the case where some transactions can gain more profits without harming other transactions.
Without considering the Victim Loss property, it is hard to validate whether such a case is a real attack since there is no victim.
To minimize the possibility of including a false attack, we require that there must be a victim who loses profits in the attack-free scenario.

$T_a^p$ is optional to perform an attack.
Eskandari et al.~\cite{eskandariSoKTransparentDishonesty2020} found that attackers may or may not need to execute another transaction after $T_v$ to collect profits\footnote{The superscript \textit{p} in $T_a^p$ means \textit{\ul{p}rofit collection}.} (Section~\ref{sec:background:attacks}).
If $\langle T_a, T_v \rangle$ already satisfies the above attack properties, it is considered as an attack without $T_a^p$.
\noindent\textbf{Discussion}:
\revision{
	Our attack model is designed to capture tuples of transactions indicating that the underlying smart contracts contain front-running vulnerabilities, i.e., allow some users front-run other users' transactions to make profits.
	Tuples of transactions identified by our attack model serve as poofs of concepts~\cite{wangReveryProofofConceptExploitable2018} for front-running vulnerabilities in the underlying smart contracts.
	In other words, transactions identified by our attack model demonstrate the feasibility of attacking $T_v$ by inserting $T_a$ and $T_a^p$ before and after $T_v$ to obtain financial profits.
	In blockchain history, the submitter of $T_a$ may not deliberately manipulate the transaction orders to make profits, though $T_a$ is accidentally executed before $T_v$, causing financial loss to the victim.
	However, $T_a$ is still a valid attack to $T_v$ and will be identified by our attack model.
	What is captured by our attack model is the de-facto attack based on the transaction execution result instead of the intention of the attackers.
}

Our attack model may also miss some attacks.
The purpose of attackers may not be to obtain financial profits as captured by our two attack properties.
However, such incentives are hard to validate.
Therefore, we limit our scope to attacks that make financial profits.
In addition, attackers may also insert multiple transactions before and after $T_v$ to perform an attack, but there is no evidence that such a scenario is common.
In the measurement study conducted by Torres et al.~\cite{torresFrontrunnerJonesRaiders2021}, only 0.025\% of front-running attacks they found involved multiple attack transactions before and after $T_v$.
This is because blockchain users need to pay extra fees for each separate transaction.
Rational attackers will merge multiple attack transactions into an atomic one to reduce costs.
Therefore, we limit our scope to those attacks involving only one attack transaction before and after $T_v$ for the sake of scalability.
The search space will otherwise grow exponentially if we consider multiple attack transactions, while there may not be many more attacks to be mined.

\revision{
	Popular front-running attack cases well-known to the blockchain community are in line with our attack model.
	Maximum Extractable Value (MEV)~\cite{daianFlashBoysFrontrunning2019} is a well-known concept in the Ethereum community, which refers to the digital assets that can be withdrawn from contracts permissionlessly.
	Assets cumulatively worth more than 696M US dollars have been withdrawn by March 2023~\cite{flashbotsMEVExplore2023a}.
	Daian et al.~\cite{daianFlashBoysFrontrunning2019} have shown that MEV withdrawal transactions are generally subject to front-running attacks, and the attacks prevalently occur on the blockchain.
	Any blockchain user is qualified to withdraw the MEV.
	Only the first of many transactions withdrawing the same MEV will successfully obtain the MEV as profits, while others will end up with no profits but still paying transaction fees.
	Therefore, different users compete for the same MEV, and the first executed transaction is attacking others for grabbing the limited number of MEV as profits.
	The front-running attacks on MEV transactions satisfy our attack properties, and they can be captured by our attack model.
}

\subsection{Attack Mining}

Existing measurement studies attempt to mine attacks using predefined patterns of transaction data or execution traces to characterize attacks~\cite{torresFrontrunnerJonesRaiders2021,daianFlashBoysFrontrunning2019}.
For instance, they consider transactions that copy the data of another transaction as attacks or search for transactions swapping tokens in the same way as described in Fig.~\ref{lst:uniswap} in a limited number of already known vulnerable token exchange contracts.
As proposed below, we do not rely on any predefined patterns and mine attacks comprehensively in the transaction history by enumerating all possible transaction combinations and identifying attacks based on our attack model.
\begin{algorithm}[t]
	\caption{Mine attacks in transaction history.}
	\label{alg:search}
	\footnotesize
	\SetKwInOut{Input}{Input}
	\SetKwInOut{Output}{Output}
	\ResetInOut{output}%
	\SetKwFunction{satisfyProperties}{satisfyProperties}
	\SetKwFunction{shouldPrune}{shouldPrune}
	\SetKwFunction{getTransactionAtIndex}{getTransactionAtIndex}
	\SetKw{Continue}{continue}
	\SetKwData{History}{$\mathbb{T}$}
	\SetKwData{Attacks}{$\mathbb{A}$}
	\SetKwData{Ta}{$T_a$}
	\SetKwData{Tv}{$T_v$}
	\SetKwData{Tp}{$T_a^p$}
	\SetKwData{ia}{$i_a$}
	\SetKwData{iv}{$i_v$}
	\SetKwData{ip}{$i_p$}

	\Input{a sequence of transactions executed in history \History}
	\Output{a set of historical attacks \Attacks}
	\BlankLine

	\Attacks $\leftarrow$ $\varnothing$\;
	\For{\ia $\leftarrow$ 0 \KwTo $|\History| - 1$ \label{alg:search:attack-loop}}{
		\Ta $\leftarrow$ \getTransactionAtIndex{\History, \ia}\;
		\For{\iv $\leftarrow$ $\ia + 1$ \KwTo $|\History| - 1$ \label{alg:search:victim-loop}}{
			\Tv $\leftarrow$ \getTransactionAtIndex{\History, \iv}\;
			\lIf{\shouldPrune{\Ta, \Tv}}{\Continue}
			\If{\satisfyProperties{\Ta, \Tv}}{
				\Attacks $\leftarrow$ $\langle \Ta, \Tv \rangle$\;
				\Continue\;
			}
			\For{\ip $\leftarrow$ $\iv + 1$ \KwTo $|\History| - 1$ \label{alg:search:profit-loop}}{
				\Tp $\leftarrow$ \getTransactionAtIndex{\History, \ip}\;
				\lIf{\shouldPrune{\Ta, \Tv, \Tp}}{\Continue}
				\If{\satisfyProperties{\Ta, \Tv, \Tp}}{
					\Attacks $\leftarrow$ $\langle \Ta, \Tv, \Tp \rangle$\;
					\Continue\;
				}
			}
		}
	}
\end{algorithm}

Algorithm~\ref{alg:search} shows the attack mining procedure in a transaction history, which is represented as a sequence of transactions $\mathbb{T}$.
It searches for the combinations of historical transactions that satisfy the attack model.
The key idea of the mining algorithm is that a successful front-running attack must result in a transaction sequence in the transaction history matching the attack scenario ($T_a \rightarrow T_v \rightarrow T_a^p$).
We can then simulate its corresponding attack-free scenario ($T_v \rightarrow T_a \rightarrow T_a^p$) to validate whether the transaction sequence satisfies our two attack properties defined in Section~\ref{sec:search:model}.
We consider every transaction in the history as a potential $T_a$ (line~\ref{alg:search:attack-loop}) and then search for any subsequent transaction $T_v$ (line~\ref{alg:search:victim-loop}) that was successfully attacked by $T_a$.
Function \texttt{satisfyProperties} executes the given transactions in the attack and attack-free scenarios and checks whether the execution result satisfies the two properties.
As explained, an attack can be accomplished by two or three transactions.
If the attack properties based on the two execution scenarios can be satisfied by $T_a$ and $T_v$, it is an attack by two transactions. %
Otherwise, the algorithm continues to search in subsequent transactions for the third transaction $T_a^p$ (line~\ref{alg:search:profit-loop}) such that $\langle T_a, T_v, T_a^p \rangle$ forms an attack.

In \texttt{satisfyProperties} function, we consider the transaction sender of $T_a$ and $T_v$ as the attacker and victim, respectively. %
Given that many attackers use bot contracts~\cite{torresFrontrunnerJonesRaiders2021} to perform attacks, we also consider the first contract that $T_a$ invokes as the attacker.
Financial gain or loss is determined using the difference in the amount of digital assets (Ether, ERC20~\cite{vogelstellerEIP20TokenStandard2023}, ERC721~\cite{entrikenEIP721NonFungibleToken2023}, ERC777~\cite{dafflonEIP777TokenStandard2023}, and ERC1155~\cite{radomskiEIP1155MultiToken2023}) that the attacker or victim receives in two transaction execution scenarios.

When checking whether the transactions satisfy the properties, we consider the existing execution in the blockchain history as the attack scenario and aggregate the total profits of attackers and victims.
For the attack-free scenario, we simulate the execution of $T_v$ based on the blockchain world state in history before $T_a$.
Then, $T_a$ and $T_a^p$ are simulated after $T_v$.
Profits of attackers and victims in the attack-free scenario are collected from the execution result of this simulation.

\revision{
	Executing transactions in attack and attack-free scenarios is expensive, especially when we aim to enumerate all possible combinations of transactions to mine attacks.
	The worst case time complexity of Algorithm~\ref{alg:search} is $O(t|\mathbb{T}|^3)$, where $t$ is the average execution time of a transaction.
}
We make improvements to the algorithm efficiency without missing attacks.
We verify the necessary conditions of the attack properties in the \texttt{shouldPrune} function and prune the search space early if the conditions are not satisfied without missing any attacks.
The primary necessary condition is that $T_a$ and $T_v$ must have read-write conflicts~\cite{luUnderstandingDetectingExposing2008} on some shared data in the blockchain world state~\cite{sergeyConcurrentPerspectiveSmart2017}, i.e.,~the account balance, contract code, and contract storage.
Inferring from the execution trace in the blockchain history, we consider $T_a$ and $T_v$ have read-write conflicts if $T_a$ modifies the same shared data that $T_v$ performs a def-clear~\cite{suSurveyDataFlowTesting2017} read.
Def-clear reads refer to those read operations the variables read by which are not written previously in the execution trace of the same transaction.
Otherwise, the execution outcome of $T_a$ and $T_v$ is irrelevant to the order between them, and the attack properties will not be satisfied.
In addition, we also prune the search space if $T_a$ and $T_v$ are submitted by the same account.

\subsection{Front-Running Attack Dataset}
\label{sec:search:dataset}

We use our attack mining algorithm to mine front-running attacks in the block range \dsFromBlock-\dsToBlock, which are the latest \dsTotalBlocks blocks when this study is conducted.
We split the entire blockchain history into windows of three blocks and slide the window with an offset of one block.
In the range of history that we are about to analyze, there are 799,998 block windows.
In each window, transactions in the consecutive blocks are concatenated into one sequence and we mine attacks in this sequence with Algorithm~\ref{alg:search}.
We analyze 16 block windows in parallel, and the mining timeout in each block window is 60s.
The mining is performed on a CentOS 8 machine with an AMD Ryzen 3975WX CPU and 512GB RAM.
It takes in total 69.54 days to finish the mining in all block windows, 7.51s for each block window on average.
We do not mine attacks in the entire blockchain history because the contracts exploited by older attacks may no longer be active.
Although the average mining time of each block window is only around half of the average Ethereum block interval (15s) \cite{etherscan.ioEthereumAverageBlock2023}, it is also impractical to mine the entire Ethereum history.
In the end, we obtain the dataset \datasetA, comprising \dsTotalAttacks attacks, from the attack mining.

\subsection{Attack Mining Evaluation}
\label{sec:search:evaluation}

To ensure the quality of dataset \datasetA, we evaluate our attack mining algorithm by answering \ref{rq1}:
\begin{itemize}[leftmargin=*]
	\item \textbf{\namedlabel{rq1}{RQ1}}: Is our mining algorithm effective in finding attacks?
	      \begin{itemize}[leftmargin=*]
		      \item \textbf{\namedlabel{rq1:precision}{RQ1-1}}: Can our algorithm effectively find real attacks?
		      \item \textbf{\namedlabel{rq1:oracle}{RQ1-2}}: Can our attack model effectively characterize attacks?
		      \item \textbf{\namedlabel{rq1:recall}{RQ1-3}}: Can our algorithm outperform the state-of-the-art attack mining technique in finding attacks?
	      \end{itemize}
\end{itemize}

\begin{table}[]
	\centering
	\label{tab:search:recall}
	\caption{Number of attacks in baseline dataset that can be found by our attack mining algorithm.}
	\begin{tabular}{l|rrr}
		\hline
		         & Displacement & Insertion & Suppression \\
		\hline
		Baseline & 2,983        & 196,691   & 50          \\
		Ours     & 2,910        & 177,222   & 0           \\
		\hline
	\end{tabular}
\end{table}

\noindent\textbf{Methodology}:
To evaluate the algorithm's precision in \ref{rq1:precision}, we manually analyze \rqOneTotalSamples attacks (\datasetS), which are randomly sampled among all attacks (\datasetA) to achieve 95\% confidence level and 5\% confidence interval.
To facilitate manual analysis, we only sample those attacks whose invoked smart contracts have source code available.
Three authors individually analyze the execution traces of each transaction in each sampled attack, interpret the semantics of underlying smart contracts, and check whether each attack found by our algorithm is an actual front-running attack according to the attack definition~\cite{nccgroupDASPTOP102021,smartcontractsecuritySWCRegistrySmart2021}, which states that attackers leverage the knowledge of future transactions to make profits.
If the three authors have different opinions, which cannot be solved after discussions, we will consider the attack as a false positive.
In the manual check, we check whether the attacker leverages the knowledge of the future victim transaction by checking whether the attacker can still obtain profits if no victim transaction is submitted.
If the attacker can no longer get profits, then it means the attacker does leverage the knowledge of the future transaction and we will consider the attack as a real front-running attack.
For instance, in Fig.~\ref{lst:transfer_manager_contract}, if the victim transaction that reveals the user's signature is not submitted, the attack transaction will have no way to pass a valid signature argument to the \texttt{relayOperation} function.
Similarly, in Fig.~\ref{lst:uniswap}, if the victim transaction that swaps \texttt{token0} to \texttt{token1} is never submitted, the exchange rate will not change after the attack transaction, and the attacker will not make profits from the price difference.
To answer \ref{rq1:oracle} and \ref{rq1:recall}, we consider the measurement study conducted by Torres~\cite{torresFrontrunnerJonesRaiders2021} as the baseline, which proposes an approach mining historical attacks using predefined transaction patterns for displacement, insertion, and suppression attacks, respectively (Section~\ref{sec:review:attacks}).
Baseline offers a dataset of three categories of attacks, as shown in the first row of Table~\ref{tab:search:recall}.
To answer \ref{rq1:oracle}, we apply our attack algorithm to mine all the attacks in the baseline dataset and check if our model can capture those attacks.
For \ref{rq1:recall}, we apply our algorithm to mine attacks in the latest 1,000 blocks (block number 11,299,000-11,300,000, containing 175,552 transactions) that the baseline mined and check whether our algorithm can find more attacks.

\noindent\textbf{Results}:
For \ref{rq1:precision}, there are only five falsely reported attacks, giving 98.69\% precision.
All of them are caused by inappropriate attack-free scenario execution.
In blockchain history, there could be many other transactions between $T_a$, $T_v$, and $T_a^p$.
When we change the transaction orders to mimic attack-free scenarios, the relative orders between $T_a$ (or $T_v$) and other transactions are also changed.
Financial profits of the attack or victim could be affected by such relative orders.
As a result, the financial profits in the attack-free scenario could be incorrectly calculated,
and false-positively reported attacks may be induced, but our manual check shows that such cases are rare.

Table~\ref{tab:search:recall} shows the experiment results for \ref{rq1:oracle}.
Out of the total 199,724 attacks in the baseline, our attack model can identify 90.19\% (180,125), indicating the generality of our attack model.
We further investigate the reasons for missed attacks.
Among the three types of front-running attacks collected by the baseline, all the suppression attacks involve multiple attack transactions before the victim transaction, which do not fit our attack model.
This is not a significant flaw in our attack model since suppression attacks only comprise a tiny portion (0.03\%) of all attacks.
We sample 61 out of 73 and 377 out of 19,469 (95\% confidence level, 5\% confidence interval) missed displacement attacks and insertion attacks to analyze the reason, respectively.
We find that 215 attacks are missed because our model is more conservative and stricter than the patterns used by the baseline.
For instance, two transactions compete to buy the same NFT~\cite{baoNonFungibleTokenSystematic2022} with ERC20 tokens, and only one of them will succeed.
The baseline considers such a case as an attack.
However, it is unknown whether the NFT is worth more than the paid ERC20 tokens, so our model does not consider it an attack.
In 160 cases, the attacker obtains zero profits or loses profits in the attack scenario.
For 19 missed attacks, we cannot re-execute the transactions in the attack-free scenario due to a violation of blockchain protocol (e.g., transaction nonce, block limit, etc.).
Thus our algorithm does not report these attacks.
The rest 44 missed cases are caused by the inappropriate attack-free scenario execution as described in the previous paragraph.

In the experiment for \ref{rq1:recall}, the baseline is able to find 277 attacks in the block range, while our algorithm is able to find 6,765 attacks, 24.42x more.
All the attacks found by the baseline can be found by our algorithm.
This result shows that our algorithm has a much higher recall rate in finding attacks.
This is because our algorithm comprehensively enumerates transactions in the blockchain history instead of relying on the heuristic patterns like the baseline.

\begin{tcolorbox}[left=0pt,right=0pt,top=0pt,bottom=0pt, enlarge top by=0.5em]
	\textit{Answer to \ref{rq1}:} Our attack mining algorithm can effectively find 24.42x more attacks than those by baseline with 98.69\% precision.
	The effectiveness of our mining algorithm ensures the quality of dataset \datasetA, which serves as a basis for the following study.
\end{tcolorbox}

\section{Vulnerability Localization and Benchmark}
\label{sec:locate}

While Section~\ref{sec:search:dataset} describes the construction of the dataset \datasetA for front-running attacks, %
the dataset cannot be used directly to evaluate various techniques' performance in front-running vulnerability detection.
Each entry in \datasetA is an attack consisting of two or three transactions but it does not pinpoint the vulnerable code snippet(s), which provide essential information to validate if vulnerabilities are correctly detected.
In this section, we present our approach to localizing the vulnerable code snippets from the transactions.

\subsection{General Ideas in a Nutshell}
\label{sec:locate:nutshell}

Pinpointing the vulnerable code snippet(s) responsible for an attack is an open problem.
In many cases, it could be the overall logic design of the vulnerable contract instead of a single line of code or a function.
For instance, in Fig.~\ref{lst:uniswap}, it is %
the algorithm design, which calculates the exchange rate of tokens, that enables front-running transactions.
None of the functions alone is vulnerable without considering the logic of others.
In this example, the attack transaction $T_a$ calls the \texttt{swap} function (Line~\ref{lst:uniswap:swap}) before the victim transaction $T_v$, reducing the amount of swapped tokens obtained by $T_v$. %
A naive approach is to consider all the code in Fig.~\ref{lst:uniswap} executed by $T_v$ in an attack scenario to be vulnerable.
However, this approach is too coarse and may falsely consider a large portion of code as vulnerable.
The code at Line~\ref{lst:uniswap:cut-fee}-\ref{lst:uniswap:fee} to pay a constant swap fee, and the body of function \texttt{logSwap} invoked at Line~\ref{lst:uniswap:log} are falsely marked vulnerable, although they are unrelated to the vulnerable logic to compute the amount of swapped tokens.

This motivates us to devise a more accurate mechanism that can scale to the large dataset \datasetA to localize vulnerable code.
In a nutshell, our approach identifies the blockchain data accessed by the victim transaction $T_v$ but altered by the attack transaction $T_a$ (\textit{attack altered data}), and performs a dynamic taint analysis~\cite{schwartzAllYouEver2010} with $T_v$ using attack altered data as taint sources.
We consider taint sinks the program location where profits earned by the victim are directly affected.
We extract the taint flow trace from source to sink and consider the contract code executed along this trace as vulnerable.
Specifically, the attack altered data of an attack is defined as follows:
\begin{definition}[Attack Altered Data]
	The attack altered data in an attack $A = \langle T_a, T_v, T_a^p \rangle$ is the blockchain data that $T_v$ performs a def-clear read after the data has been stored by $T_a$ in the attack scenario.
\end{definition}

In Fig~\ref{lst:uniswap}, both $T_a$ and $T_v$ invokes function \texttt{swap}.
$T_a$ modifies contract variables at line~\ref{lst:uniswap:alteration}, which are later loaded by $T_v$ at line~\ref{lst:uniswap:calculation}.
We consider these two variables (\texttt{reserve0} and \texttt{reserve1}) as taint sources in the dynamic taint analysis of $T_v$.
Profits earned by the victim are transferred at line~\ref{lst:uniswap:profit}, whose amount is decreased because of the attack.
We thus consider line~\ref{lst:uniswap:profit} as the taint sink.
We then compute the vulnerable code snippet by extracting the flow from source to sink, i.e.,~line~$\ref{lst:uniswap:calculation}\rightarrow\ref{lst:uniswap:external-call}\rightarrow\ref{lst:uniswap:profit}$.
The vulnerable logic that computes the token exchange rates using attack altered data is identified, while contract code at line~\ref{lst:uniswap:cut-fee}-\ref{lst:uniswap:fee} and \texttt{logSwap} function are excluded.
Compared with the naive approach, which marks all lines of code, we only mark three lines in function \texttt{swap} and \texttt{doSwap} as vulnerable.

\subsection{Localize Vulnerability with Influence Trace}

Now we present how to mechanically localize vulnerable code snippet(s) from an attack $A=\langle T_a, T_v, T_a^p\rangle$.
First, we localize the taint sources by identifying attack altered data in the attack scenario.
Blockchain shared data, i.e.,~account balances, contract code, and contract storage, which are modified in $T_a$ and read without preceding writes (def-clear~\cite{suSurveyDataFlowTesting2017} reads) in $T_v$, is considered as attack altered data.
Those operations in $T_v$ that perform def-clear reads on attack shared data are considered taint sources.
Second, we localize the taint sink that is held responsible for the loss of victim's financial profits.
We conduct a manual analysis on the same set of attack samples \datasetS as in Section~\ref{sec:search:evaluation} and check how victims' financial profits are influenced by attack altered data.
We make an interesting finding that \textit{all attacks can be summarized into three attack patterns} based on how the attack altered data influences victim transactions, namely Path Condition Alteration, Computation Alteration, and Gas Estimation Griefing.
Taint sinks are defined accordingly for different attack patterns.

\textit{Path Condition Alteration}:
\begin{lstlisting}[language=Solidity]
if (altered(sharedData)) {
    uint profit = computeProfit();
    victim.transfer(profit);
}
\end{lstlisting}
The above code snippet shows the first attack pattern.
The victim's profit depends on a path condition evaluated using attack altered data.
The example shown in Fig.~\ref{lst:transfer_manager_contract} falls into this pattern.
In this pattern, the root cause is that the path condition is manipulatable by attackers, while the computation of profits is not.
We consider the conditional statement as the taint sink.
Note that we cannot use the profit transfer operations as taint sinks since they do not necessarily data-depend on the attack altered data.

\textit{Computation Alteration}:
\begin{lstlisting}[language=Solidity]
uint profit = calculateProfits(
  altered(sharedData)
);
victim.transfer(profit);
\end{lstlisting}
The above code snippet shows the computation alteration pattern.
The computation of the victim's financial profit is manipulated without changing the execution path.
Attacks on the example exchange contract in Fig.~\ref{lst:uniswap} falls into this pattern.
We consider the statement that transfers profits to the victim as the taint sink.

\textit{Gas Estimation Griefing}:
\begin{lstlisting}[language=Solidity]
parameterizedExpensiveOperation(
  altered(sharedData)
);
victim.transfer(profit);
\end{lstlisting}
Gas estimation griefing is different from the previous two patterns.
Instead of manipulating the execution path or computation outputs, the attacker attacks by leveraging the gas model of Ethereum.%
Blockchain users need to estimate and specify a sufficient gas limit before submitting transactions, otherwise the execution fails.
The gas consumption of transactions may depend on the attack altered data, in which case attackers can make the actual gas consumed by the victim transaction larger than the user-specified limit.
As a result, attackers could make victim transaction fail to their own benefit.
Note that the underlying smart contracts may not contain vulnerabilities because the attack will not succeed if the victim transactions are equipped with sufficient gas.
Therefore, we do not define taint sink or localize vulnerabilities for gas estimation griefing attacks.
We classify the attack $A$ into attack patterns by inspecting the execution traces of $T_v$ in the two execution scenarios, and identify the taint sink $\delta$ accordingly.
Let $\tau$ and $\tau^f$ denote the two execution traces of $T_v$ in the attack and attack-free scenarios, respectively.
If $\tau$ throws an out-of-gas exception while $\tau^f$ does not, $A$ is considered a gas estimation griefing attack and excluded from our vulnerability localization. %
To classify the attack into the other two patterns,
we first extract the sequences of program locations performing digital asset transfers in $\tau$ and $\tau^f$, and denote them as $[\tau_0,\tau_1,...,\tau_p]$ and $[\tau^f_0,\tau^f_1,...,\tau^f_q]$, respectively.
We distinguish the attack patterns of an attack by checking the proper prefix of $\tau$ and $\tau^f$.

Case 1 (Path condition alteration): $\exists i, 0 \leq i \leq \max(p,q)$ such that $\tau_i \neq \tau^f_i$ and $\forall{j}, 0 \leq j < i \land \tau_j = \tau^f_j$.
We categorize attack $A$ as a path condition alteration attack, and consider the first divergence point between $\tau$ and $\tau^f$ as $\delta$ for this attack, where $\tau_i$ and $\tau^f_i$ control-depend on $\delta$. %

Case 2 (Computation alteration): $\forall i, 0 \leq i \leq \max(p,q) \land \tau_i = \tau^f_i$.
We categorize attack $A$ as a computation alteration attack.
Note that there must exist $j$, $0 \leq j \leq \max(p,q)$, such that the transfer operation at program location $\tau_j$ (or $\tau^f_j$) transfers a different amount of digital assets.
We consider the program location $\tau_j$ as $\delta$ for this attack.

Finally, we extract the flow trace from the taint source to the sink.
We call this flow trace \textit{influence trace}, covering the code that depends on attack altered data and influences the victim's financial profits.
Note that for one attack, there can be multiple taint sources and thus multiple influence traces since the attack altered data may be loaded as tainted values in different places.
We use influence traces to over-approximate the vulnerability location of an attack by considering all code executed in an influence trace as vulnerable.
It is a trade-off between localizing a smaller range of vulnerable code and ensuring the vulnerability is covered by the marked code,
because it is hard to precisely and correctly localize without contract specifications from developers.

\subsection{Vulnerability Localization Evaluation}
\label{sec:locate:evaluation}

We evaluate the effectiveness of our vulnerability localization approach with the following research question:
\begin{itemize}
	\item \textbf{\namedlabel{rq3}{RQ2}}: Can our approach precisely identify the exploited vulnerabilities from real-world attacks?
\end{itemize}

\noindent\textbf{Methodology}:
We answer \ref{rq3} from two aspects with the dataset \datasetS as mentioned in Section~\ref{sec:search:evaluation}.
First, we check whether the exploited vulnerability can be localized by our approach, given a real-world attack.
We perform a manual analysis on each attack in \datasetS.
The five falsely reported attacks identified previously are excluded.
As pointed out by previous studies~\cite{luuMakingSmartContracts2016,kolluriExploitingLawsOrder2019,tsankovSecurifyPracticalSecurity2018,boseSAILFISHVettingSmart2022,eskandariSoKTransparentDishonesty2020,sergeyConcurrentPerspectiveSmart2017,wangDetectingNondeterministicPayment2019} and SWC Registry~\cite{smartcontractsecuritySWCRegistrySmart2021}, the root cause of front-running vulnerability is the race condition~\cite{commonweaknessenumerationCWECWE362Concurrent2023} in the smart contract where the execution result depends on the order of transactions.
Specifically, in the manual check, we consider vulnerability locations to be the code locations where the victim transaction loads the attack altered data modified by the attack transaction, i.e.,~locations of the read-write conflicts between the victim and attack transactions.
We do not consider locations of write-write conflicts as vulnerability locations since in these cases, the victim transaction's execution result is not affected.
If the code snippet identified by our approach includes the vulnerability locations, we consider our vulnerability localization result as a true positive.
Three authors individually check for each attack, and all disagreements are discussed until they are resolved.
Second, we check whether our approach can precisely pinpoint vulnerable code without including many unrelated code.
We build a baseline based on the naive approach mentioned in Section~\ref{sec:locate:nutshell}, i.e., considering all code executed by $T_v$ as vulnerable.
We collect and compare the number of EVM~\cite{woodEthereumSecureDecentralised2020} instructions identified as vulnerable code by the baseline and our approach, respectively.
We measure how many unrelated code our approach can reduce compared to the baseline.

\begin{figure}
	\centering
	\includegraphics[scale=0.42]{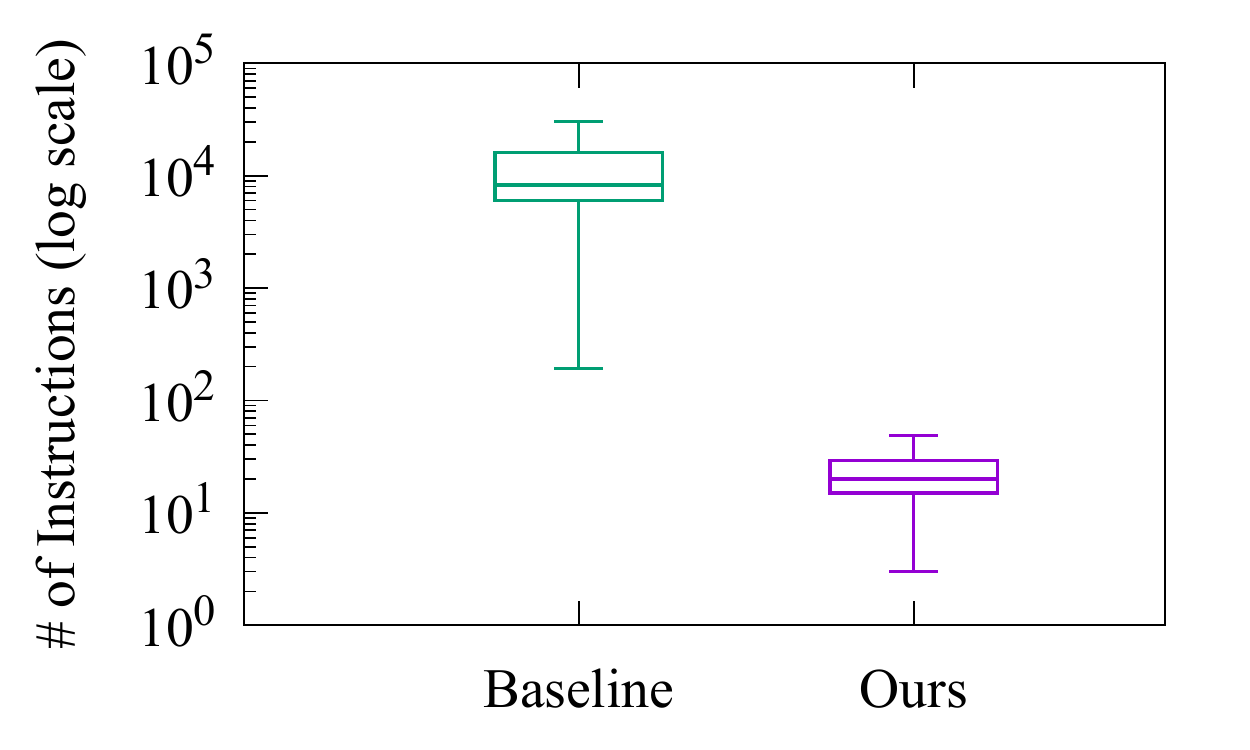}
	\caption{Distribution of the number of EVM instructions marked as vulnerable by the baseline and our approach for each attack in \datasetS.}
	\label{fig:reduction}
\end{figure}
\begin{figure}
	\centering
	\includegraphics[scale=0.42]{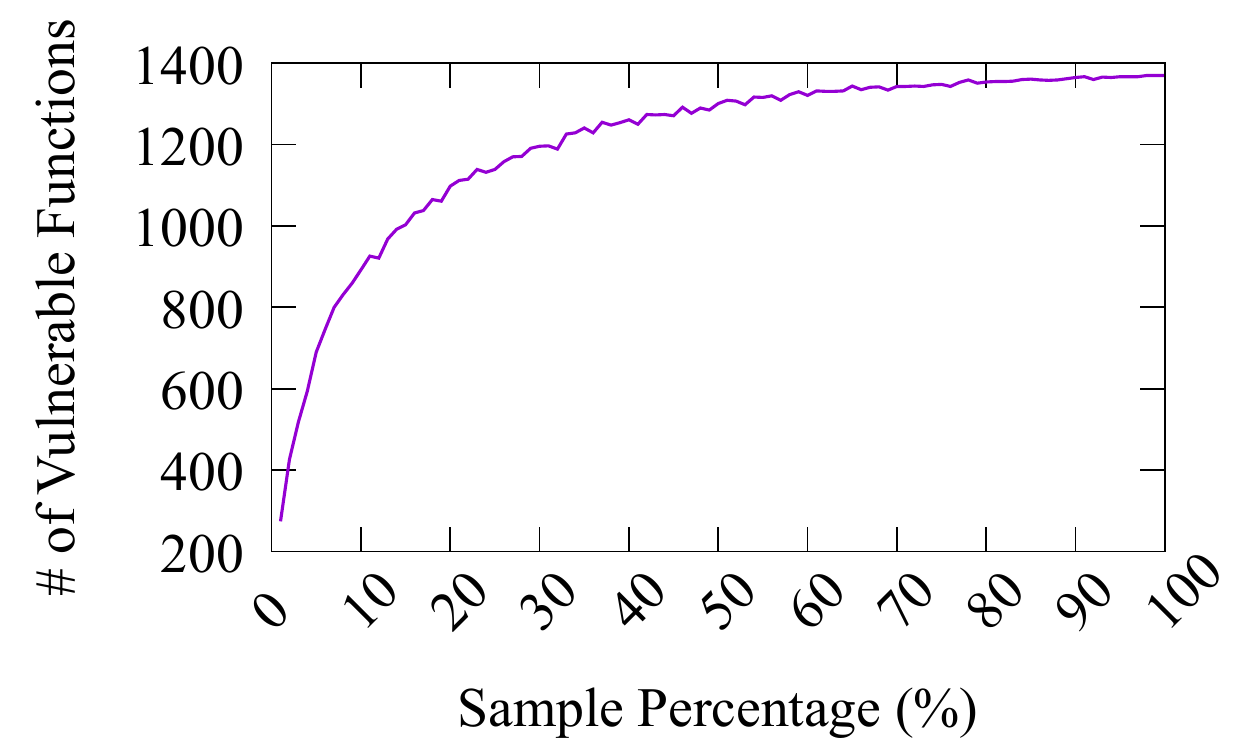}
	\caption{The total number of distinct vulnerable functions in top-1200 contracts saturates as more attacks are sampled from \datasetP.}
	\label{fig:vulnerability-saturation}
\end{figure}

\noindent\textbf{Results}:
In our manual inspection, we find that the identified vulnerable code is able to cover the vulnerable logic exploited in all 378 attacks of \datasetS. %
As shown in Fig.~\ref{fig:reduction}, on average, 25.25 EVM instructions are marked vulnerable for each attack.
Compared to the baseline, our approach marks only 0.34\% of those instructions marked by the baseline as vulnerable, resulting in a 99.66\% reduction rate.
One can leverage our approach to construct effective and large benchmarks on front-running vulnerabilities in the absence of contract specifications. %

\begin{tcolorbox}[left=0pt,right=0pt,top=0pt,bottom=0pt, enlarge top by=0.5em]
	\textit{Answer to \ref{rq3}:} Our localization approach is effective in pinpointing vulnerable code to a much smaller range than the baseline without missing any exploited vulnerabilities.
\end{tcolorbox}

\subsection{Benchmark Construction}
\label{sec:locate:benchmark}

To build a benchmark for the comparison of vulnerability detection tools, we extract influence traces for each attack in dataset \datasetA.
Attacks that result in multiple influence traces are excluded to avoid ambiguities in vulnerability localization.
We mark all public contract functions that are executed in the influence traces as vulnerable.
We label vulnerable functions because the contract analyzers evaluated in Section~\ref{sec:evaluation} commonly report problems at the function level.

We do not have the ground truth of all vulnerabilities in each contract. %
For each contract included in the benchmark, we are unsure if our benchmark has labeled most of the vulnerabilities ever exploited in blockchain history, because we did not mine the entire blockchain history in Section~\ref{sec:search:dataset}.
To mitigate this threat, we focus on a set of most popularly attacked contracts and check if additional vulnerabilities in these contracts can be labeled when more attacks in the blockchain history are considered.

The following strategy is adopted for benchmark construction.
We select top-$N$ popularly attacked contracts and only consider vulnerable functions in these contracts in our benchmark.
The popularity is measured by the invocation frequency of each contract in all the influence traces of all attacks in dataset \datasetA.
From \datasetA, we select a subset $\datasetP$ of attacks whose influence traces only involve contracts in these top-$N$ contracts.
Then, $n\%$ attacks are sampled from \datasetP.
We increase $n$ from $1$ to $100$ with step $1$ and compute the number of distinct vulnerable functions localized in each sample.
If the total number of distinct vulnerable functions saturates as $n$ increases, it indicates that we are unlikely to find new vulnerable functions in these top-$N$ contracts even if we keep mining for more attacks in the blockchain history.
In other words, the occurrence of saturation hints that exploited vulnerable functions in the selected contracts have been mostly labeled. Intuitively, we want to include more contracts in the benchmark while saturation is observed. %
In our study, we set $N$ to 1200.
Fig.~\ref{fig:vulnerability-saturation} shows the total number of distinct vulnerable functions against the sampling size of \datasetP.
The number of vulnerable functions only increases by 0.36\% (from 1,365 to 1,370) between 90\% and 100\% samples.
The saturation would gradually disappear when the number of contracts considered increases beyond 1,200.

\begin{figure}
	\centering

\end{figure}

Therefore, we build benchmark \benchmarkA on the 1,200 selected contracts by analyzing the attacks in dataset \datasetP.
Vulnerable functions in these contracts are labeled with influence traces, as previously explained for each attack.
In addition, we also use influence traces to remove those attacks that exploited the same vulnerability.
If multiple attacks have the same influence trace, we consider that they are duplicate exploitation and include only one of them.
To facilitate manual analysis, we include only those attacks occurring at the functions whose source code is available on Etherscan~\cite{etherscanEthereumBlockchainExplorer2021}.
As a result, we construct the benchmark \benchmarkA consisting of \benchmarkTotalAttacks attacks with vulnerable functions localized in \benchmarkTotalContracts distinct contracts.

\section{Evaluation of Existing Tools}
\label{sec:evaluation}

In this section, we demonstrate the use of \benchmarkA to understand the status quo of front-running vulnerability detection.
We evaluate tools that implement state-of-the-art vulnerability detection techniques and answer the following research question.
\begin{itemize}[leftmargin=*]
	\item \textbf{\namedlabel{rq4}{RQ3}}: How many vulnerabilities can existing tools detect in our benchmark?
	\item \textbf{\namedlabel{rq5}{RQ4}}: What are the limitations of existing tools in detecting front-running vulnerabilities?
\end{itemize}

\subsection{Tool Selection}
We conduct a systematic literature review to collect tools that implement representative state-of-the-art smart contract analysis techniques detecting front-running vulnerabilities.
Based on the guideline from Brereton et al.~\cite{breretonLessonsApplyingSystematic2007}, we search for related publications in top-tier conferences and journals, perform a backward snowballing to find more literature, and collect available tools from them.
To largely include the state-of-the-art tools, we use \textit{contract}, \textit{ethereum} as search keywords and search for publications in all CORE~\cite{coreincCORERankingsPortal2023} A/A* ranked venues in software engineering and security fields with research code: 4612, 4604, and 0803.
For each matching publication, we read the abstract and apply the following criteria:
1) Empirical studies and literature reviews are excluded.
2) Only papers about detecting contract vulnerability without requiring additional information from developers are included.
At this step, we are able to collect 47 publications in 18 venues.
We continue to perform a backward snowballing by searching for related literature in the references of these publications.
In the end, we find 17 additional papers, technical reports, and GitHub repositories.
From these literature, we collect available tools, which implement the techniques that support the detection front-running vulnerabilities.
In the end, we collect \totalTools tools suitable for our empirical evaluation, namely Oyente~\cite{luuMakingSmartContracts2016}, Securify~\cite{tsankovSecurifyPracticalSecurity2018},  Ethracer~\cite{kolluriExploitingLawsOrder2019}, Mythril~\cite{muellerSmashingEthereumSmart2018}, Conkas~\cite{velosoConkas2022}, Securify2~\cite{srilabSecurify2SecurifyMaster2023}, and Sailfish~\cite{boseSAILFISHVettingSmart2022}.
The techniques used in these tools are discussed in Section~\ref{sec:review:techniques}.

\begin{table}[]
	\centering
	\small
	\resizebox{0.85\linewidth}{!}{%
		\begin{threeparttable}
			\caption{Vulnerability detection result of each tool on benchmark \benchmarkA.}
			\label{tab:detection-result}
			\begin{tabular}{c|rrr|rrr}
				\hline
				\multirow{2}{*}{Tool} & \multicolumn{3}{c|}{Attacks} & \multicolumn{3}{c}{Contracts\tnote{1}}                                             \\
				                      & TP                           & FN                                     & Recall & N/A\tnote{2} & Timeout & Failure \\
				\hline
				Oyente                & 0                            & 513                                    & 0\%    & 0            & 0       & 0       \\
				Mythril               & 16                           & 497                                    & 3.12\% & 0            & 0       & 20      \\
				Conkas                & 0                            & 513                                    & 0\%    & 0            & 4       & 205     \\
				Securify              & 31                           & 482                                    & 6.04\% & 0            & 0       & 69      \\
				Ethracer              & 13                           & 500                                    & 2.53\% & 0            & 1       & 4       \\
				Securify2             & 0                            & 513                                    & 0\%    & 23           & 0       & 206     \\
				Sailfish              & 8                            & 505                                    & 1.56\% & 23           & 1       & 186     \\
				\hline
			\end{tabular}
			\begin{tablenotes}
				\footnotesize
				\item [1] There are in total \benchmarkTotalContracts distinct contracts involved in all influence traces of attacks in \benchmarkA. One distinct contract may be involved in influence traces of multiple attacks.
				\item [2] The contract is not compilable for tools that analyze bytecode, or not flattenable for tools that analyze source code in single file.
			\end{tablenotes}
		\end{threeparttable}
	}
\end{table}

\subsection{Experiment Design}
\label{sec:evaluation:design}

For each attack in benchmark \benchmarkA, we run experiments to check whether the exploited vulnerability can be detected by each tool.
We use each tool to analyze all contracts whose code is marked vulnerable in \benchmarkA.
Note that none of the selected tools support analyzing a group of contracts together, so we let each tool analyze contracts individually.
Two tools, i.e.,~Securify2 and Sailfish, can only analyze contracts in source code in a single file.
We use Hardhat~\cite{nomicfoundationHardhatEthereumDevelopment2023} toolchain to flatten contract source code into a single file and let these two tools analyze the flattened source file.
For all other tools that analyze contract bytecode, we compile the contract source code into Byzantium EVM bytecode~\cite{ethereumfoundationHistoryForksEthereum2023}, which is the most compatible version supported by all tools.
Different tools may detect various types of vulnerabilities.
However, we are only interested in the result of front-running vulnerability, i.e.,~event ordering bugs in Ethracer, state inconsistency bugs in Sailfish, and transaction order dependency in all other tools.

We set the analysis timeout of each tool equally to three hours, which is larger than the longest timeout among the evaluation experiments of these tools' original papers.
With benchmark \benchmarkA, we adopt the following approach to check whether a vulnerability exploited by an attack is detected by each tool.
In the detection results of one tool, we consider one attack is \textit{true positive (TP)} if the tool reports problems in any of the vulnerable functions localized with this attack as described in Section~\ref{sec:locate:benchmark}.
If none of these functions is reported vulnerable by the tool, we consider the attack is \textit{false negative (FN)}.
The recall rate of each tool is computed with the total number of TP attacks divided by the total number of attacks in \benchmarkA.
Note that our benchmark does not label vulnerable functions that have not been exploited in the blockchain history.
If one tool reports problems in other functions outside our benchmark, we cannot conclude whether they are false alarms or not.
Thus, we do not evaluate the precision of these tools.

\subsection{Evaluation Results}

Table~\ref{tab:detection-result} shows the vulnerability detection result of each tool.
On the left side, we report the number of TP and FN attacks for each tool using the criteria mentioned in Section~\ref{sec:evaluation:design}.
For all tools, the number of missed vulnerabilities is significant.
The best tool, Securify, only has a 6.04\% recall rate.
The majority of vulnerabilities are missed by all tools.
Our evaluation shows the poor performance of state-of-the-art tools with a large-scale benchmark.
A similar conclusion was drawn by the previous study~\cite{durieuxEmpiricalReviewAutomated2020} with a small benchmark of four contracts, which are not representative since the average lines of code for each contract is only 33.75, and none of them is a real-world contract used on the blockchain.
In comparison, our benchmark contains much more representative vulnerable contracts and can better reveal the real performance of vulnerability detection techniques.

We also found that several tools could not successfully analyze many contracts, as shown on the right side of Table~\ref{tab:detection-result}.
Some tools timeout on the analysis of a few complex contracts, as shown in the Timeout column.
Securify2 and Sailfish work on Solidity source code and can only analyze contracts written in a single file.
The source code of 116 out of 235 contracts in our benchmark spreads across multiple files.
Although we try to flatten multi-file contracts into a single file, there are 23 contracts that cannot be flattened due to cyclic dependencies between source files, as shown in the N/A column.
In addition, we also found that Securify2 and Sailfish have poor support for contracts written in newer Solidity versions, resulting in a large amount of analysis failure.
We found that other bytecode analyzers, especially Conkas, crash on a large portion of contracts.
Similar crashes are encountered by other users according to the tools' issue tracker and they have not been fixed by developers.

\begin{tcolorbox}[left=0pt,right=0pt,top=0pt,bottom=0pt, enlarge top by=0.5em]
	Answer to \ref{rq4}:
	Existing tools detect at most 6.04\% of vulnerabilities in \benchmarkA, suggesting their weaknesses in exposing front-running vulnerabilities in real-world contracts.
	Effective detection tools are urgently needed.
\end{tcolorbox}

\begin{table}
	\centering
	\resizebox{\linewidth}{!}{%
		\begin{threeparttable}
			\caption{Manual analysis results for the limitations of each tool.}
			\label{tab:limitation}
			\begin{tabular}{c|rr|rrrrr}
				\hline
				\multirow{3}{*}{Tool} & \multicolumn{2}{c|}{FN Attacks} & \multicolumn{5}{c}{Limitation}                                                                                                                        \\
				                      & \multirow{2}{*}{Total}          & \multirow{2}{*}{Sample}        & \multicolumn{2}{c}{Code Analysis} & \multicolumn{2}{c}{Oracle} & \multirow{2}{*}{Unknown\tnote{2}}                   \\
				                      &                                 &                                & IC\tnote{1}                       & CS\tnote{1}                & P\tnote{1}                        & T\tnote{1} &    \\
				\hline
				Oyente                & 390                             & 194                            & 124                               & -                          & 65                                & 5          & 0  \\
				Mythril               & 370                             & 189                            & 132                               & -                          & 52                                & 5          & 0  \\
				Conkas                & 12                              & 12                             & 7                                 & -                          & 5                                 & 0          & 0  \\
				Securify              & 155                             & 155                            & 0                                 & -                          & 0                                 & 155        & 0  \\
				Ethracer              & 491                             & 216                            & 133                               & 31                         & 19                                & 0          & 33 \\
				Securify2             & 3                               & 3                              & 0                                 & -                          & 0                                 & 3          & 0  \\
				Sailfish              & 47                              & 47                             & 41                                & -                          & 0                                 & 6          & 0  \\
				\hline
			\end{tabular}
			\begin{tablenotes}
				\item [1] IC, CS, P, and T stand for Lack Support for \underline{I}nter-\underline{C}ontract Analysis, \underline{C}onstraint \underline{S}olving for Cryptographic Operations, Weak Detection \underline{P}attern, and Lack of \underline{T}oken Support, respectively.
				\item [2] We were unable to identify the limitations resulting in 33 FN attacks for Ethracer.
			\end{tablenotes}
		\end{threeparttable}
	}
\end{table}

\subsection{Discussion on Limitations of Existing Techniques}

We randomly sample FN attacks for each tool with 95\% confidence level and 5\% confidence interval and manually analyze them to understand the reasons behind the poor performance of existing techniques.
We focus only on those FN attacks whose concerned contracts can all be successfully analyzed by the tool since we aim to investigate the limitations of each tool's technique rather than its implementation.
The large second column of Table~\ref{tab:limitation} shows the number of sampled attacks.

Existing tools commonly detect vulnerabilities in two phases, namely code analysis and oracle checking.
First, bytecode or source code is analyzed to extract the semantics of contracts such as control flow and data flow using symbolic execution or static analysis.
The extracted semantics are then examined against predefined vulnerability oracles to detect the existence of vulnerabilities.
A tool can exhibit limitations in any of the two phases.
Table~\ref{tab:limitation} shows the number of attacks whose vulnerabilities were missed by each tool due to the limitations in each vulnerability detection phase.
Note that tools may miss some vulnerabilities due to limitations in both phases.
For such cases, we followed the order of tools' working procedures and categorized them into the code analysis phase.
For the other limitations that cannot be categorized into the two phases, we put them into the \textit{Unknown} column.

\subsubsection{Two Limitations in Code Analysis} %
In the code analysis phase, we found two common limitations of existing tools: lack of support for inter-contract analysis and unavailability of efficient constraint solvers for non-linear computation such as cryptographic and hashing operations.
Column IC and CS in Table~\ref{tab:limitation} present the number of attacks whose vulnerabilities are missed due to each limitation.

\noindent\textbf{Inter-Contract Analysis}:
We find that existing techniques lack support for inter-contract analysis of the scenarios where a contract invokes another contract during its execution.
Existing techniques are designed to analyze contracts individually, while ignoring their possible interactions with other contracts.
For example, the vulnerability in Fig~\ref{lst:uniswap} cannot be detected if each contract is individually analyzed because the vulnerable exchange rate computation (line~\ref{lst:uniswap:calculation}) and the loading of attack altered data (\texttt{reserve0} and \texttt{reserve1}) reside in different contracts.
In tools based on symbolic execution, i.e.,~Oyente, Mythril, Conkas, and Ethracer, the execution details of the external contract code are omitted  (e.g.,~line~\ref{lst:uniswap:external-call} in Fig.~\ref{lst:uniswap}).
In addition, their symbolic execution does not properly handle the return value of external calls.
The symbolic execution using the return value will halt because the EVM opcode \texttt{RETURNDATACOPY} is not properly defined in their implementation.
Similarly, Sailfish also omits external contract invocations during its interpretation of contract semantics.
As a result, contract semantics are not fully analyzed by these tools.
To show the impact of this limitation, we manually inspected the attacks in dataset \datasetS and found that in 222 out of 383 sampled attacks, the influence traces span across multiple contracts, i.e.,~external contract calls are involved in exploiting the vulnerabilities.
This result shows that the lack of inter-contract analysis support can lead to many undetected vulnerabilities.

\revision{
	Supporting inter-contract analysis is, however, challenging.
	The address of external contracts being invoked is usually not statically decidable.
	The address may be stored in a contract storage variable set by other users or provided as input by transactions.
	In other words, any contract on the blockchain can be a potential callee during analysis.
	Worse still, each potential callee contract may have further external contract invocations, making the size of the contract code to analyze grow exponentially.
	Future techniques should adopt effective strategies to prune irrelevant code so that the detection is scalable to find inter-contract vulnerabilities.
}

Securify and Securify2, however, adopt an over-approximation approach by simply considering that every external contract call is manipulatable by attackers and the return value is malicious.
Thus they are not subject to this limitation.
Nonetheless, as shown in the study from Durieux et al.~\cite{durieuxEmpiricalReviewAutomated2020}, such over-approximation can induce many false alarms, which, however, is out of the scope of our study.
\begin{tcolorbox}[left=0pt,right=0pt,top=0pt,bottom=0pt]
	\begin{finding}
		Existing tools cannot properly handle external contract calls to perform precise inter-contract analysis.
		Such limitations make them miss the detection of many vulnerabilities.
		Future detection techniques of front-running vulnerabilities should include inter-contract analysis.
	\end{finding}
\end{tcolorbox}

\noindent\textbf{Constraint Solving}:
In addition to inter-contract analysis, Ethracer also suffers from the limitation of constraint solving in the code analysis phase.
Ethracer uses dynamic symbolic execution to generate concrete transactions invoking functions and covering as many execution paths as possible for the contract under analysis.
Such concrete transactions are executed in different orders to trigger the front-running vulnerabilities in a fuzzing process.
SMT solver is used to resolve function inputs.
However, it is impossible to solve constraints involving cryptographic operations.
The path condition at line~\ref{lst:transfer_manager_contract:verify} in Fig~\ref{lst:transfer_manager_contract} involves digital signature verification.
It is impossible for techniques like that of Ethracer to resolve a valid input to satisfy this path condition using existing SMT solvers.
Another unsolvable operation is the keccak256 hash operation, which is widely used in smart contracts, such as computing the address of values in mapping or array variables.
If such types of variables are used in path conditions, Etheracer will also fail to generate concrete transactions to cover some or all execution paths and thus cannot detect the vulnerabilities in those uncovered paths.
Other symbolic execution-based tools do not suffer from this limitation since they do not need to generate concrete inputs for functions.
A common workaround solution for the cryptographic and hashing operations in constraints is to replace their results with new intermediate symbolic values.

\revision{
	For techniques that need to generate concrete transactions, it may not be feasible to generate inputs that satisfy various path conditions involving cryptographic operations.
	However, the large transaction history can be leveraged to tackle the issue.
	In the transaction history, there may be some transaction inputs that can satisfy the various constraints that an SMT solver can hardly resolve.
	Analyzers can mutate from those inputs to generate concrete transactions to test other contracts.
}

\begin{tcolorbox}[left=0pt,right=0pt,top=0pt,bottom=0pt]
	\begin{finding}
		The widely used cryptographic operations in smart contracts make it hard for an SMT solver to generate concrete inputs, weakening the capability of existing tools in exploring transaction executions for attack scenarios.
		Future studies should seek better approaches to effectively generate concrete transactions that satisfy path conditions with cryptographic and hashing operations.
	\end{finding}
\end{tcolorbox}

\subsubsection{Two Limitations in Oracle}

Limitations of detection oracles occur in two ways.
First, some tools fail to define proper vulnerability patterns for effective vulnerability detection.
Second, many tools only check whether the vulnerability can cause a loss of Ethers while lacking the support of digital assets in token standards (e.g.,~ERC20, ERC721, ERC777, and ERC1155).
As a result, they fail to detect many vulnerabilities that cause the loss of these tokens.
Table~\ref{tab:limitation} presents the number of attacks whose vulnerabilities are missed due to each limitation.

\noindent\textbf{Patterns}:
Each technique defines specific patterns to identify vulnerabilities in smart contracts.
Oyente computes the number of digital assets transferred with symbolic execution.
Oyente reports front-running vulnerabilities if there exist two execution paths transferring in different symbolic amounts.
However, vulnerable contracts of computation alteration attacks like Fig.~\ref{lst:uniswap} do not fall into this pattern since the amount of digital assets that the victim obtains in the attack and attack-free scenarios are symbolically the same.
In addition, some path condition alteration attacks may also not fall into the pattern if there is no digital asset transfer in some execution paths, e.g.,~Fig.~\ref{lst:transfer_manager_contract}.

Mythril and Conkas identify vulnerabilities by checking if the receiver or amount of digital asset transfers depends on shared variables modifiable by other transactions.
However, some path condition alteration attacks like Fig.~\ref{lst:transfer_manager_contract} may be missed since the profit transfer control-depends, instead of data-depends, on the attack altered data.

Ethracer checks whether different invocation orders of two different functions result in different blockchain world states after the execution.
However, it does not consider failed victim transactions as attack consequences.
Vulnerabilities are not reported if one of the functions throws exceptions in one of the execution orders.
In fact, attacks on contracts may result in failed victim transactions, e.g.,~Fig.~\ref{lst:transfer_manager_contract}.
Thus, Ethracer will also miss this vulnerability.

Securify, Securify2, and Sailfish use a general pattern, checking whether digital asset transfers depend on blockchain-shared data through either control flow or data flow.
This generic pattern can conceptually capture all vulnerabilities in our benchmark.
However, these tools lack the support of tokens, as discussed below, and thus miss most of the vulnerabilities that result in the loss of tokens.
Another limitation of adopting such a general pattern is that a lot of false alarms in the reported problems.
Durieux et al.~\cite{durieuxEmpiricalReviewAutomated2020} has shown the high false positive rate of Securify.
This is because the dependency on global variables does not necessarily result in profitable opportunities that attackers can exploit.

\revision{
	The two attack properties proposed in our attack model can also serve as a general oracle in front-running vulnerability detection techniques.
	In other words, contract analyzers can check if there exists a tuple of transactions invoking the contract under analysis, such that the two attack properties are satisfied.
	One major challenge to adopt this oracle is the large search space for analyzers.
	Oracles used by existing tools, except for Ethracer, are based on the execution of a single transaction.
	However, checking our attack properties requires the ordered execution result of multiple transactions.
	The complexity of analysis will grow exponentially as the size of contract code increases.
	Novel techniques to reduce the search space during analysis are desired.
}

\begin{tcolorbox}[left=0pt,right=0pt,top=0pt,bottom=0pt]
	\begin{finding}
		The detection patterns of many tools are weak in capturing front-running vulnerabilities in real-world smart contracts.
		Vulnerability detection tools should be updated with patterns based on real-world vulnerable contracts.
	\end{finding}
\end{tcolorbox}

\begin{table}
	\centering
	\caption{The number of attacks in which each type of victim's financial profits decreases.}
	\label{tab:evaluation:profit}
	\resizebox{\linewidth}{!}{%
		\begin{tabular}{r|rrrrr}
			\hline
			                     & Ether   & ERC20   & ERC721 & ERC777 & ERC1155 \\
			\hline
			Attacks in \datasetA & 118,702 & 184,987 & 2,931  & 1,060  & 537     \\
			\hline
		\end{tabular}
	}
\end{table}

\noindent\textbf{Token Support}:
In addition to detection patterns, the negligence of profit making in tokens by existing techniques causes many attacks undetected.
All tools except Ethracer support only Ether as digital assets in pattern matching of vulnerability detection.
As a result, the vulnerabilities of many attacks where victims lose tokens instead of Ethers are missed.
In contracts of Fig.~\ref{lst:uniswap} and Fig.~\ref{lst:transfer_manager_contract}, attackers and victims obtain financial profits in terms of ERC20 tokens instead of Ethers.
Tools not supporting ERC20 are not aware of such attack profits and thus do not report vulnerabilities.
Note that in Table~\ref{tab:limitation}, we analyzed the limitation of detection patterns supposing the tool supports Ethers and all tokens.
We concluded that a vulnerability is missed due to the token support limitation only when the pattern can capture the vulnerability and the tool does not support tokens.
The reason is that it is easy to extend the support for tokens.
To identify a token transfer, tool developers only have to check whether the data of an external call matches the transfer function signature of the standard token interface.
We considered token support as the least impacting limitation.
To show the impact of this limitation, we collect the total number of attacks in which each type of financial profit is involved in the attack, as shown in Table~\ref{tab:evaluation:profit}.
Note that each attack can have more than one type of financial profit, i.e.,~there can be overlaps between different types in Table~\ref{tab:evaluation:profit}.
Tokens are even more prevalent than Ether in the dataset presented in Section~\ref{sec:search:dataset}.
The other three types of tokens also have a non-negligible share in the profits of front-running attacks.
The support of profit analysis in tokens is essential to vulnerability detection for smart contracts.

\begin{tcolorbox}[left=0pt,right=0pt,top=0pt,bottom=0pt]
	\begin{finding}
		Many front-running vulnerabilities are missed by existing detection tools due to their negligence of profit making in tokens.
		The support of profit analysis in tokens is essential to detect front-running vulnerabilities in real-world smart contracts.
	\end{finding}
\end{tcolorbox}

\subsection{Other Vulnerabilities Identified by Existing Tools}

\revision{
	Given the poor performance of existing tools on our front-running vulnerability benchmark, one further question is whether existing tools can detect other front-running vulnerabilities that are missed by our benchmark.
	Therefore, we inspect the detection results of all \totalTools tools on all \benchmarkTotalContracts contracts in our benchmark \benchmarkA.
}

\revision{
	As mentioned before, each tool has named vulnerabilities resulting in front-running attacks in different ways.
	We only focus on vulnerability reports related to front-running attacks, i.e., the event ordering bugs in Ethracer, state inconsistency bugs in Sailfish, and transaction order dependency bugs in all other tools.
	The \totalTools tools in total report a total of 293 front-running vulnerable functions, which is much more than the number of vulnerabilities they can catch in our benchmark as shown in Table~\ref{tab:detection-result}.
	However, as pointed out by previous studies, existing tools have a high false positive rate~\cite{durieuxEmpiricalReviewAutomated2020,perezSmartContractVulnerabilities2021}.
	It is unknown whether these 293 reported vulnerabilities are true positives or not.
	We do not have ground truth on the reported vulnerable functions, so we follow a similar practice of a previous study~\cite{durieuxEmpiricalReviewAutomated2020} and consider a function is truly vulnerable if at least two tools raise the alarm for this function.
	The rationale behind this is that: a vulnerability is more likely a true positive if several tools have an agreement on it.
}

\revision{
	As a result, there are only 24 functions that are reported as vulnerable by at least two tools, indicating that the majority of reported vulnerabilities are likely to be false positives.
	This result aligns with previous studies~\cite{durieuxEmpiricalReviewAutomated2020,perezSmartContractVulnerabilities2021}, which evaluate the detection precision of existing tools.
	Of the 24 vulnerable functions, one is included in our benchmark, while others are not.
	We investigate all of the rest 23 functions and find that they are all false positives.
	There are four main reasons that these functions are not vulnerable.
	Eight functions are false positives since they are \textit{view} functions in smart contracts, which cannot transfer any assets or modify any blockchain state.
	Eight functions are false positives since they can only be called by one specific account (e.g.,~admin or owner).
	Malicious attackers are not allowed to invoke the functions.
	Four functions are false positives since transactions invoking these functions can only access or update the state space belonging to the transaction submitter.
	Different transactions have no shared data so it is impossible to launch front-running attacks on another transaction submitted by different users.
	The rest three functions are false positives since transactions invoking these functions can obtain more profits if front-run by other transactions according to the semantics of the contracts.
	We do not consider them exploitable vulnerabilities since victims do not suffer from loss.
	Therefore, our investigation shows that existing tools can hardly identify issues that are not included in our benchmark.
	They may report many vulnerabilities, but the majority of them are false positives.
}

\revision{
	Nevertheless, it is possible that some vulnerabilities detected by existing tools are missed by our benchmark because we do not search the entire blockchain history, and some vulnerabilities may never be exploited in history.
	As mentioned in Section~\ref{sec:locate:benchmark}, for vulnerabilities exploited in the not-searched history, the number of them in our selected contracts is likely small.
	For vulnerabilities never exploited, our study gives a high priority to the soundness of our benchmark (i.e.,~included vulnerabilities are true positives).
	Therefore, we only consider exploited vulnerabilities as ground truth, and those unexploited ones may be missed.
}

\section{Threats to Validity}

A validity threat in our study is that our analysis is based on the attacks in \dsTotalBlocks blocks instead of the entire blockchain history.
We mitigate this threat by using the latest blocks to improve the representativeness of the attacks in our benchmark.
We also focus on 1200 popularly attacked contracts, as discussed in Section~\ref{sec:locate:benchmark}, and show that most exploited vulnerabilities in these contracts have been identified in our benchmark.
In addition, we may execute existing contract analyzers improperly.
We mitigate this threat by strictly following the instructions and actively communicating with the tool authors when encountering issues.
\revision{
	Another validity threat arises from the subjectivity in manual analysis when evaluating our attack mining algorithm and influence trace.
	We mitigate this threat by setting objective criteria for the ground truth of true front-running attacks and real vulnerable contract code in manual analysis (Section~\ref{sec:search:evaluation} and Section~\ref{sec:locate:evaluation}).
	In addition, the manual analysis result is obtained by a consensus among independent manual checks from three different authors, all of whom have more than three years of experience in the security analysis of smart contracts.
}

\section{Conclusion}

In this paper, we design an algorithm to automatically mine real-world front-running attacks.
We localize vulnerable contract code using dynamic taint analysis on the found attacks and build a benchmark of front-running vulnerabilities.
Based on the benchmark, we perform an empirical evaluation of \totalTools state-of-the-art vulnerability detection techniques.
We find that the performance of these techniques is still limited and identify four limitations in their code analysis process and vulnerability detection oracles.
The implementation of our approach, benchmark, and tool evaluation results are available on GitHub: \url{https://github.com/Troublor/erebus-redgiant}.

\ifCLASSOPTIONcompsoc
	\section*{Acknowledgments}
\else
	\section*{Acknowledgment}
\fi
This work is supported by
National Natural Science Foundation of China (Grant No. 61932021),
Hong Kong Research Grant Council/General Research Fund (Grant No. 16205821 and Grant No. 14206921),
Hong Kong Research Grant Council/Research Impact Fund (Grant No. R5034-18),
Natural Sciences and Engineering Research Council of Canada (Grant No. RGPIN-2022-03744),
Natural Sciences and Engineering Research Council of Canada (Grant No. DGECR-2022-00378),
and Guangdong Basic and Applied Basic Research Fund (Grant No. 2021A1515011562).

\ifCLASSOPTIONcaptionsoff
	\newpage
\fi

\bibliographystyle{IEEEtran}
\bibliography{BoSE,CommonSE,Interdisciplinary,PL,Additional}

\begin{IEEEbiography}[{\includegraphics[width=1in,height=1.25in,clip,keepaspectratio]{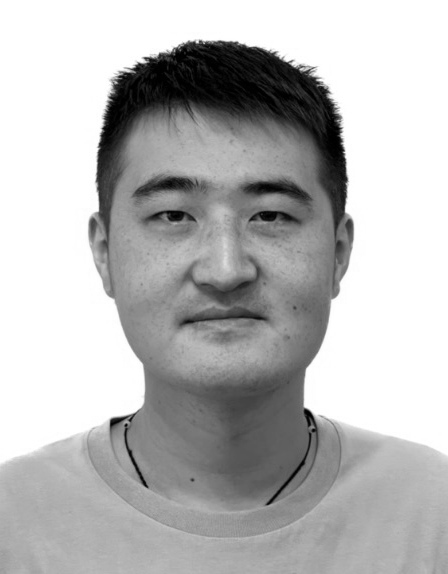}}]{Wuqi Zhang}
	received his B.Eng. degree from Northeastern University (NEU), China, in 2019.
	He is currently a Ph.D. student in the CASTLE research group at The Hong Kong University of Science and Technology (HKUST).
	His research interest includes program analysis, software testing, and security verification with a focus on blockchain applications, deep learning applications, and IoT software.
	His academic achievement has been recognized with a variety of awards, including Outstanding Graduate of Liaoning Province, HKUST RedBird Academic Excellence Award, and HKUST Overseas Research Award.
	He is also actively serving in the software engineering research community as a committee member of the Artifact Evaluation track in several venues, including ISSTA and ICSE.
	More information can be found on his personal website: \url{https://troublor.xyz}.
\end{IEEEbiography}
\begin{IEEEbiography}[{\includegraphics[width=1in,height=1.25in,clip,keepaspectratio]{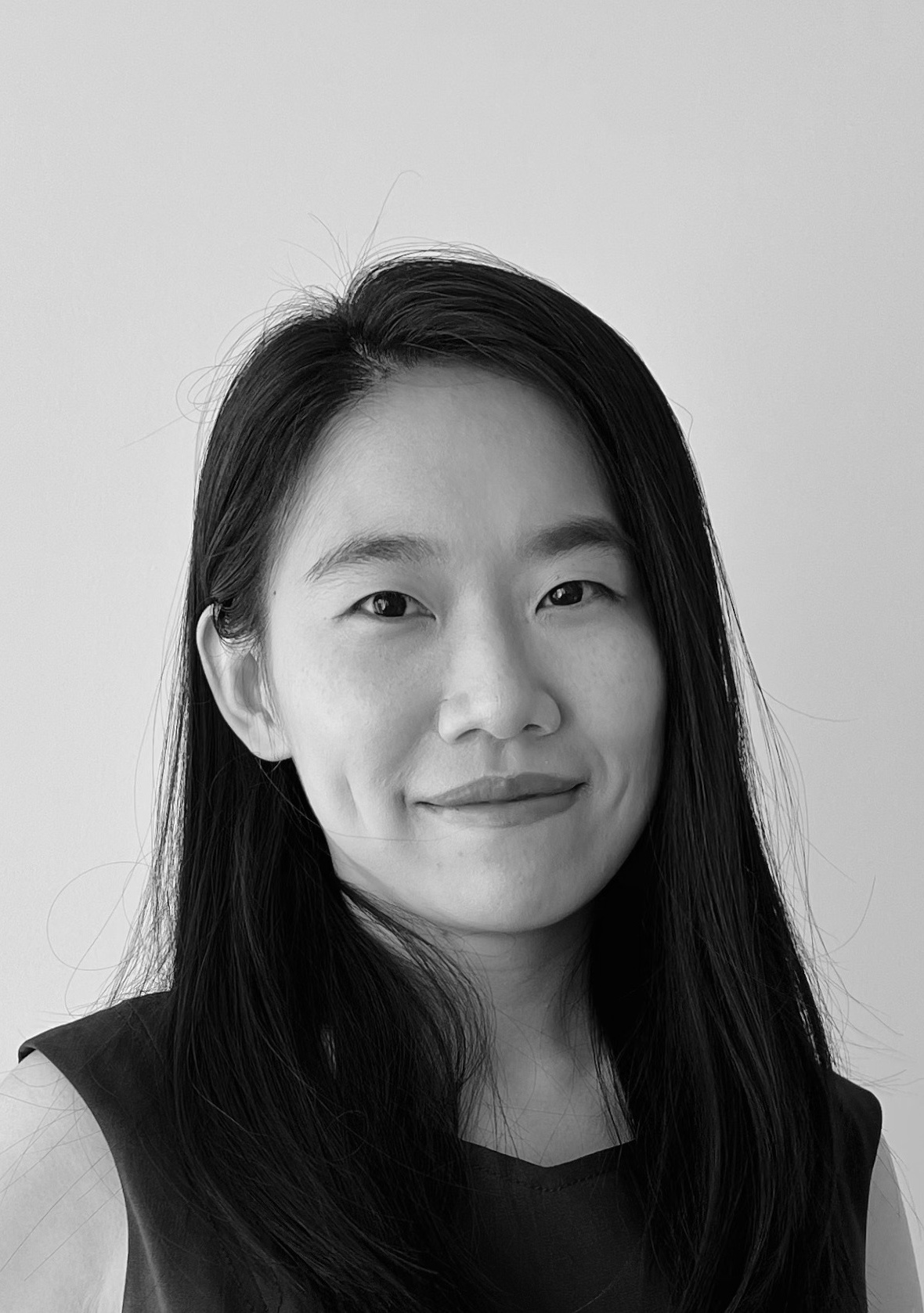}}]{Lili Wei}
	is an assistant professor at McGill University. Prior to joining McGill University, she received her Ph.D. degree and worked as a post-doctoral fellow at the Hong Kong University of Science and Technology. Her research interests lie in program analysis and testing with a focus on mobile apps, smart contracts and IoT software. Her research outcomes were recognized by several awards, including an ACM SIGSOFT Distinguished Paper Award, an ACM SIGSOFT Distinguished Artifact award, a Google PhD Fellowship and a Microsoft Research Asia PhD Fellowship. She is also actively serving the software engineering research community. She actively served on the program committees and organizing committees for a variety of major software engineering venues and co-chaired the program commitee of MOBILESoft 2023. She also received a Distinguished Reviewer Award from ASE 2022. More information can be found on her personal website: \url{https://liliweise.github.io}.
\end{IEEEbiography}
\begin{IEEEbiography}[{\includegraphics[width=1in,height=1.25in,clip,keepaspectratio]{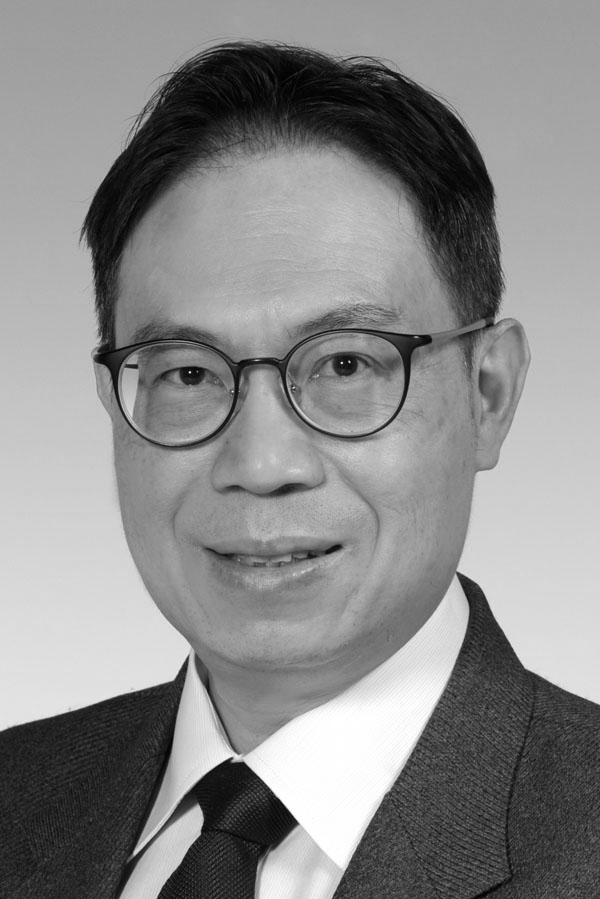}}]{Shing-Chi Cheung}
	received his doctoral degree in Computing from the Imperial College London. After that, he joined the Hong Kong University of Science and Technology (HKUST), where he is a professor of Computer Science and Engineering. He founded the CASTLE research group at HKUST and co-founded in 2006 the International Workshop on Automation of Software Testing (AST), which is now an annual IEEE international conference. He was the General Chair of the 22nd ACM SIGSOFT International Symposium on the Foundations of Software Engineering (FSE 2014). He was an editorial board member of the IEEE Transactions on Software Engineering (TSE, 2006-9). His research interests focus on the use of advanced testing, analysis, AI techniques and empirical experimentation techniques for the detection, diagnosis and repair of faults in dependable and intelligent software system. He is an ACM Distinguished Member and an IEEE Fellow. More information about his CASTLE research group can be found at \url{http://castle.cse.ust.hk/castle/people.html}.
\end{IEEEbiography}
\begin{IEEEbiography}[{\includegraphics[width=1in,height=1.25in,clip,keepaspectratio]{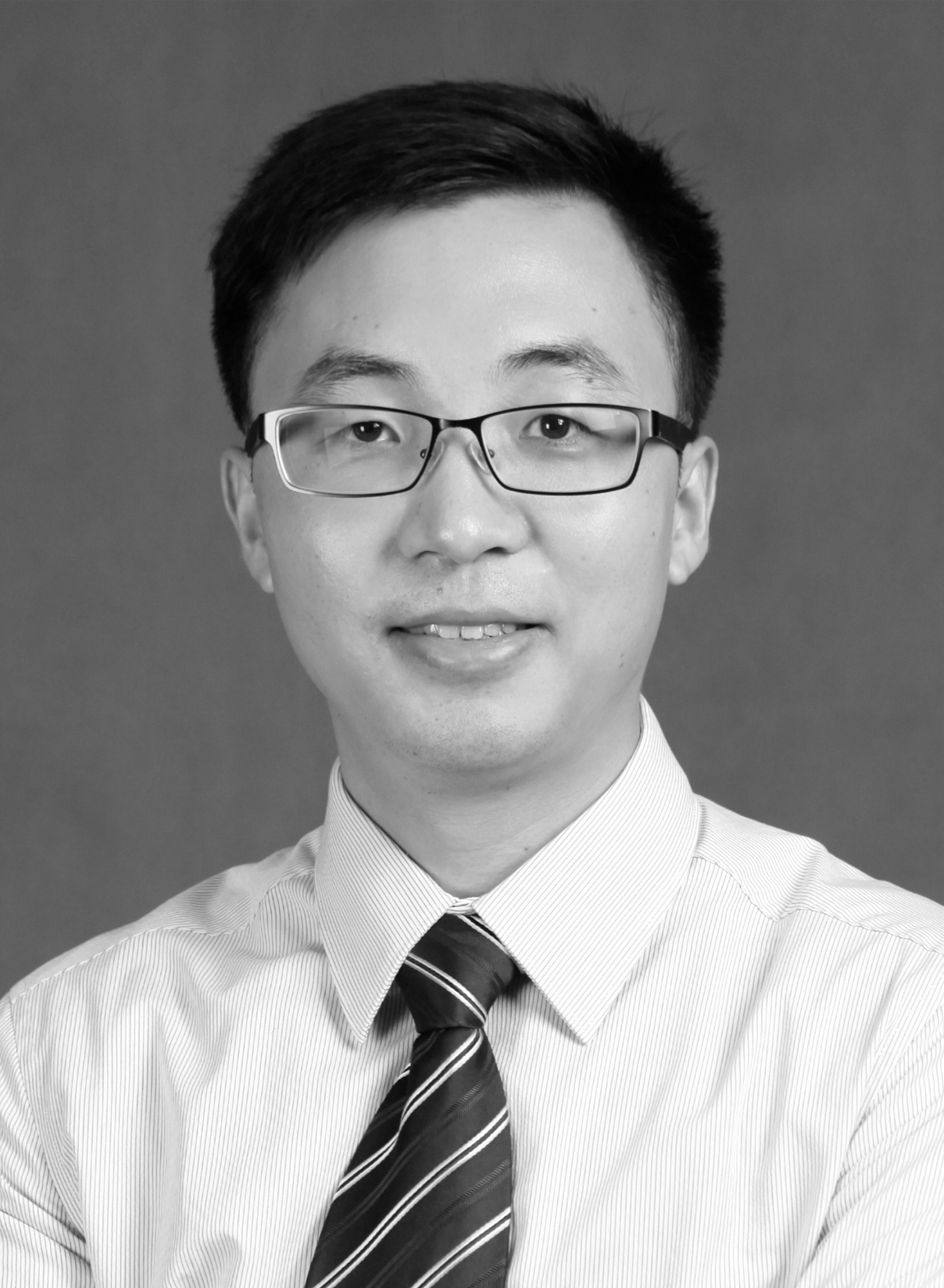}}]{Yepang Liu}
	is a tenure-track assistant professor at the Southern University of Science and Technology (SUSTech). He leads the Software Quality Lab and is the director of the Trustworthy Software Research Center of the Research Institute of Trustworthy Autonomous Systems. He obtained his Ph.D. degree from the Hong Kong University of Science and Technology (HKUST) in 2015, under the supervision of Prof. Shing-Chi Cheung, and B.Sc. degree with honor from Nanjing University in 2010. Prior to joining SUSTech, he was a post-doctoral fellow at the Cybersecurity Lab and CASTLE Lab of HKUST. His research interests include empirical software engineering, software testing and analysis, autonomous systems, mobile computing, and cybersecurity. His work has been published in many reputable software engineering venues, including TSE, TOSEM, ICSE, ESEC/FSE, ASE, and ISSTA, and has been recognized by several awards, including three ACM SIGSOFT Distinguished Paper awards and one ACM SIGSOFT Distinguished Artifact award. He also actively serves as reviewers and organizers for major international conferences and journals and received an ACM SIGSOFT Service award. More information can be found on his personal page: \url{https://yepangliu.github.io/}.
\end{IEEEbiography}
\begin{IEEEbiography}[{\includegraphics[width=1in,height=1.25in,clip,keepaspectratio]{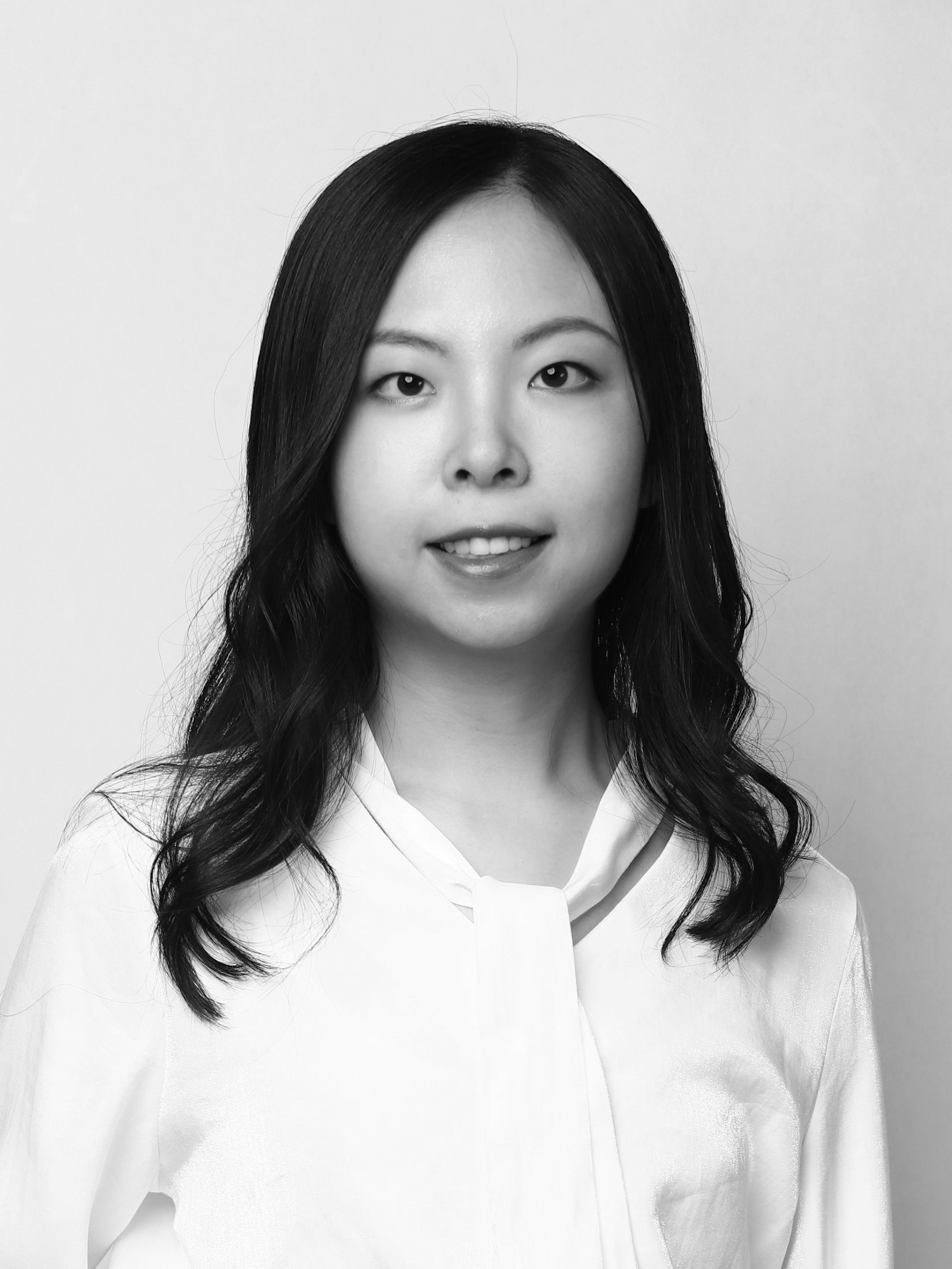}}]{Shuqing Li}
	received the B.Eng. degree from Southern University of Science and Technology (SUSTech). She is currently working toward the Ph.D. degree at the Department of Computer Science and Engineering, The Chinese University of Hong Kong (CUHK). Her research interests include software testing, software analysis, and intelligent software engineering.
	More information can be found on her personal website: \url{https://shuqing-li.github.io/}.
\end{IEEEbiography}
\begin{IEEEbiography}[{\includegraphics[width=1in,height=1.25in,clip,keepaspectratio]{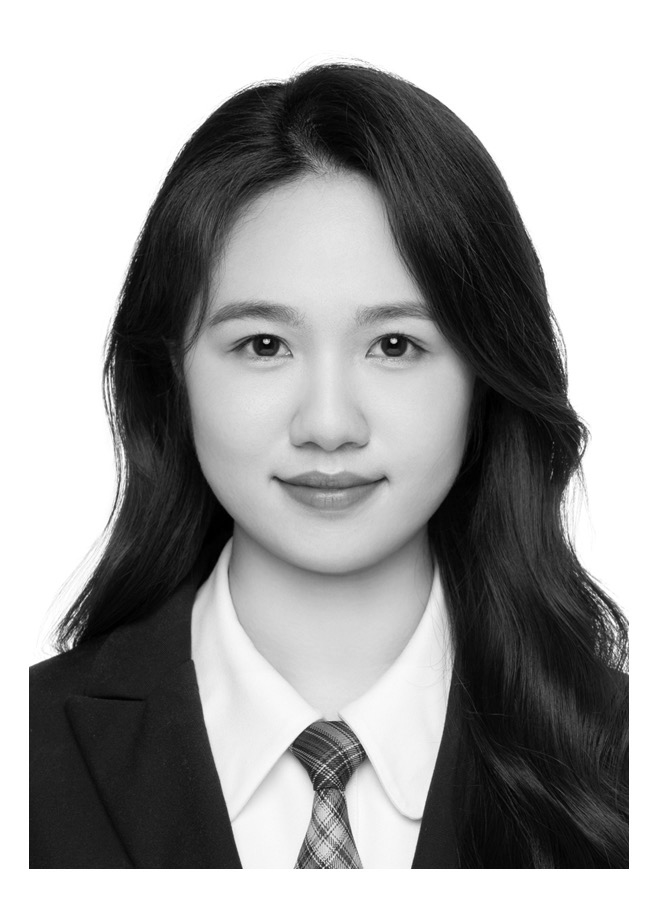}}]{Lu Liu}
	is a PhD student at the Hong Kong University of Science and Technology, majoring in Computer Science and Engineering. Prior to her PhD studies, she received a BS degree in software engineering from Wuhan University. Her research interests focus on testing and analyzing vulnerabilities in Ethereum smart contracts.
\end{IEEEbiography}
\begin{IEEEbiography}[{\includegraphics[width=1in,height=1.25in,clip,keepaspectratio]{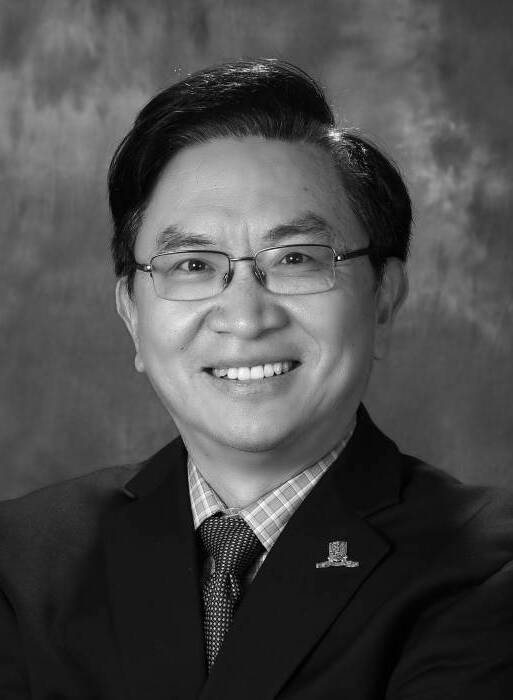}}]{Michael R. Lyu}
	received his B.S. in Electrical Engineering from National Taiwan University, Taipei, Taiwan; his M.S. in Computer Science from University of California, Santa Barbara, USA; and his Ph.D. in Computer Science from University of California, Los Angeles, USA.  He is currently Choh-Ming Li Professor of Computer Science and Engineering in The Chinese University of Hong Kong. Prof. Lyu's research interests include software engineering, software reliability, machine learning, cloud and mobile computing, and distributed systems. He has published over 600 refereed journal and conference papers in his research areas. His Google Scholar citation is over 49,000, with an h-index of 109. Prof. Lyu initiated the first International Symposium on Software Reliability Engineering (ISSRE) in 1990. He was an Associate Editor of IEEE Transactions on Reliability, IEEE Transactions on Knowledge and Data Engineering, IEEE Transactions on Services Computing, and Journal of Information Science and Engineering. He is currently on the editorial board of IEEE Access, Wiley Software Testing, Verification and Reliability Journal (STVR), and ACM Transactions on Software Engineering Methodology (TOSEM). Prof. Lyu was elected to IEEE Fellow (2004), AAAS Fellow (2007), ACM Fellow (2015), and named IEEE Reliability Society Engineer of the Year (2010). He was granted with China Computer Federation (CCF) Overseas Outstanding Contributions Award in 2018, and the 13th Guanghua Engineering Science and Technology Award in 2020. He was also named in The AI 2000 Most Influential Scholars Annual List with three appearances in 2020.
\end{IEEEbiography}

\end{document}